\documentclass[11pt]{article}
\usepackage{amsfonts}
\usepackage{amsmath}
 \usepackage{amssymb}
 \newfont{\bbbold}{msbm10}

 \def\cD{{\cal D}}

 \def\cM{{\cal M}}
 \def\cN{{\cal N}}
 \def\cO{{\cal O}}
 \def\cP{{\cal P}}
 
 \def\cR{{\cal R}}
 \def\cS{{\cal S}}
 \def\cT{{\cal T}}

 \newfont{\goth}{eufm10 scaled \magstep1}

 \def\a{\alpha}
 \def\b{\beta}
 \def\c{\gamma}\def\C{\Gamma}
 \def\d{\delta}\def\D{\Delta}

 \def\l{\lambda}
 \def\m{\mu}
 \def\n{\nu}

 \def\s{\sigma}\def\S{\Sigma}
 
 \def\th{\theta}

 \def\be{\begin{equation}}\def\ee{\end{equation}}
 \def\bea{\begin{eqnarray}}\def\eea{\end{eqnarray}}
 \def\ba{\begin{array}}\def\ea{\end{array}}
 
 \def\o{\omega}\def\O{\Omega}

 \def\una{\underline a}
 \def\unb{\underline b}

 \def\unm{\underline m}\def\unM{\underline M}
 \def\unmu{\underline \mu}
 \def\unN{\underline N}
 \def\unP{\underline P}

 \def\str{\rm str}

 \def\3dt{\dot{3}}

\def\Co{\stackrel{0}{\C}}
\def\cDo{\stackrel{0}{\cD}}
\def\oo{\stackrel{0}{\o}}

 \let\la=\label

 {}

 \def\nn{\nonumber}
 \def\bd{\begin{document}}
 \def\ed{\end{document}}
 \def\ealt{\end{alignat}} 
 \def\bead{\begin{alignat}{2}}
 \def\beat{\begin{alignat}{3}}
 \def\bea{\begin{eqnarray}}
 \def\ba{\begin{array}}\def\ea{\end{array}}
 \def\eea{\end{eqnarray}}
 \def\ft#1#2{{\textstyle{{\scriptstyle #1}\over {\scriptstyle #2}}}}
 \def\fft#1#2{{#1 \over #2}}
 \newcommand{\eq}[1]{(\ref{#1})}
 \def\eqs#1#2{(\ref{#1}-\ref{#2})}
 \def\det{{\rm det\,}}
 \def\tr{{\rm tr}}\def\Tr{{\rm Tr}}
  \def\str{{\rm str}} \def\diag{{\rm diag}}
 \def\sdet{{\rm sdet}}
\def\ealt{\end{alignat}}
 
\begin{document}

 \thispagestyle{empty}

 \hfill{G\"{o}teborg ITP preprint}

  \hfill{\today}

 \vspace{20pt}

 \begin{center}
 {\Large{\bf Curved 11D Supergeometry}}
 \vspace{30pt}

 {Dimitrios Tsimpis} \vskip .3cm {Institute for Theoretical Physics}

G\"{o}teborg University and 

Chalmers University of Technology


SE-412 96 G\"{o}teborg, Sweden
%
%

 \vspace{60pt}

 \end{center}

 {\bf Abstract} We examine the 
theta-expansion of the eleven-dimensional 
supervielbein. We outline a systematic procedure which 
can be iterated to any order. 
We give explicit expressions for the vielbein 
and three-form potential components 
up to order ${\cal O}(\th^5)$. 
Furthermore we show that at each order in the number of 
supergravity fields, in a perturbative expansion around flat 
space, it is possible to obtain exact expressions to all orders 
in theta. We give the explicit expression at linear order 
in the number of fields and we 
show how the procedure can be iterated to any desired order.
As a byproduct we obtain the complete linear coupling 
of the supermembrane to the background supergravity fields,  
covariantly in component form. We discuss the implications 
of our results for M(atrix) theory.

 {\vfill\leftline{}\vfill \vskip  10pt

 \baselineskip=15pt \pagebreak \setcounter{page}{1}


\section{Introduction}

It has been hoped that the 
quantization of the world-volume theory of the eleven-dimensional 
supermembrane \cite{bsta, bstb}, 
in analogy to the quantization of the superstring, 
would furnish a microscopic description of M-theory. 
However, a head-on approach to this problem has been stalled by 
certain features of the membrane which make it much 
less tractable than the superstring: 
the presence of 
nonlinearities and the absence of conformal invariance. 

On the other hand, it has been known for some time that supersymmetric 
matrix quantum mechanics (henceforth: matrix model) 
emerges as the finite-N regularization of the 
world-volume theory of 
the supermembrane in flat eleven-dimensional spacetime 
in the light-cone frame \cite{whn}. 
Excitement at the prospect of being able to 
describe the supermembrane by such a simple theory as ordinary 
quantum mechanics was temporarily halted by the realization 
that the spectrum of the theory was continuous \cite{wln}.  
The latter fact was initially thought to 
signal the instability of the membrane.

More recently, however, the large-N limit of the matrix model 
was conjectured to give the complete 
description of M-theory in uncompactified 
flat space in the ``infinite momentum frame'' \cite{bfss}, 
and the issue of the continuous spectrum was resolved  
by recognizing that the Hilbert space 
of the theory contains multi-particle states. 
The original BFSS conjecture\footnote{Somewhat  confusing is the fact that 
the BFSS succeeded the 
so-called ``old matrix model'', but   
the latter has recently succeeded in re-emerging as the new new one; 
as a result the BFSS is now the new old matrix model. } 
was later extended 
to finite N \cite{susskind}. According to this, 
finite-N supersymmetric quantum mechanics describes
the discretized light-cone quantization of M-theory 
with N units of compact momentum. 
It was subsequently argued that the finite-N 
conjecture follows from the BFSS \cite{sen, s}.

The argument of \cite{sen, s} is still valid in the case of  
toroidal compactifications of M-theory. 
An extension of the conjecture to include 
general curved backgrounds is much
less clear however, and several complicating issues arise 
\cite{dos, do}. The original BFSS conjecture 
was recently generalized in \cite{bmn} to include M-theory on the 
maximally-supersymmetric eleven-dimensional plane wave background. 
In \cite{tra} the linear couplings of the matrix model
to general background supergravity fields were derived 
within the context of matrix theory, up to quadratic 
order in the fermionic membrane coordinates  ($\th$).

The same problem was approached from the point of view of the 
supermembrane in \cite{dnp}, generalizing the work of \cite{ggk} 
for the superparticle. The authors of \cite{dnp} 
constructed the light-cone gauge-fixed membrane vertex operators 
describing the interactions of massless supermultiplets 
in the linearized approximation. 
These vertex operators 
are nothing but the linear 
couplings of the membrane to a general supergravity background. 
Contact  with the matrix model was made by subsequently performing a 
finite-N regularization. A covariant computation
to all orders in $\th$ was deemed unfeasible and was not 
attempted. In fact no covariant vertex operators 
had been constructed, 
even for the superstring or the superparticle, until recently 
within Berkovits' pure-spinor approach \cite{ba, bb}. 
The vertex operators can be used to compute (or rather: to define)
membrane scattering amplitudes \cite{p, pnpw}. They are therefore 
interesting in their own right, in the hope that 
they may prove useful in probing the dynamics of M-theory.

The 
(component form of the) world-volume theory of the 
supermembrane in a general supergravity background is  also 
an essential input in membrane-instanton computations 
\cite{bbs, hm, mps, lima, limab}. Such nonperturbative effects have 
recently attracted a lot of attention in the context of 
cosmological models with moving branes. (The literature on the 
subject is already vast: see 
for example \cite{ekpyrotic, pyrotechnic, kallosh} 
and papers citing those).

In principle, all information about the coupling of the 
eleven-dimensional supermembrane to the background 
supergravity fields is 
contained in the action of \cite{bsta, bstb}. The latter, however, is given 
in superspace coordinates, in terms of the background supervielbein. 
Extracting information in component form from the supermembrane 
action boils down to the problem of obtaining the 
$\th$-expansion of the supervielbein.  
For a general 
superspace geometry, this problem has so far 
only been tackled  
iteratively. Clearly, iterating all the way up to 
$\th^{32}$ (when the series terminates) is 
exceedingly tedious. 
All-order results are only known in special cases, namely 
for flat space and for 
the coset superspaces $AdS_4\times S^7$,  
$AdS_7\times S^4$ \cite{krr, dffftt, wpps, claus} 
(for a related type IIB discussion 
of $AdS_5\times S^5$ see \cite{mt}).

Partial results regarding terms to order $\th^2$ 
in the expansion of 
the eleven-dimensional supervielbein have appeared in 
the literature in \cite{plef}. The authors of 
\cite{plef} used a method called ``gauge completion'' which,  
as far as elegance is concerned, leaves a lot to be desired.
The authors of \cite{gris} computed explicitly 
up to ${\cal O}(\th^3)$ for a general background 
(and up to ${\cal O}(\th^4)$ for a bosonic background) using 
the method of normal coordinates \cite{norcor}. The latter 
is also known as the ``covariant $\th$-expansion'' --the 
terminology referring to the fact that the  gauge fields 
(graviton, gravitino) enter through the
supervielbein and superconnection, and their 
derivatives enter through
covariant field-strengths. The method of 
normal coordinates takes advantage of the 
superspace formulation \cite{cf, brh} of eleven-dimensional
supergravity \cite{cjs} 
and, as opposed to gauge completion, is unambiguous 
and systematic.

In this paper we use normal coordinates  
(we take the point of view that it is the most natural 
generalization of the Wess-Zumino gauge) to derive recursion 
relations between different levels of the $\th$-expansion.
All results are eventually expressed in component form. 
We obtain the explicit expansion of the supervielbein to order 
${\cal O}(\th^5)$ and we find agreement with the 
existing literature referred to in the previous paragraph. 
More significantly, however, our way of presenting 
the recursion relations 
makes possible the following important observation: 
at each order in the number of fields,  
in a perturbative expansion around flat space,  
{\it it is possible to obtain exact expressions to all orders in} $\th$. 
We derive the explicit expressions at linear order in the 
number of fields, in section 6.1. We also explain how 
the procedure can be iterated to higher orders.

Let us sketch the basic idea of our method. 
Our particular gauge-fixing choice leads to the 
equations \footnote{We employ standard superspace notation. 
Further details can be found in the following sections.}
\bead
\th^\m (E_\m{}^A-\d_\m{}^A)&=0\nn\\
\th^\m \O_{\m A}{}^B&=0\nn
\end{alignat}
and
\be
\th^\m\partial_\m=\th^\m\d_\m^\a\nabla_\alpha ~.\nn
\end{equation}
The above allow us to systematically translate 
the order of the $\theta$-expansion to the number of spinor 
derivatives, where 
the action of the latter on the various superfields 
is known. 
We note that knowledge of the $\th$-expansion of the superfield 
$T_{ab}{}^\a$, the covariant gravitino field-strength,  
suffices to obtain the $\th$-expansion of
all other superfields, the vielbein in particular.
On the other hand, in an expansion around flat space, 
the $n$-th level ($T^{(n)}_{~~ab}{}^\a$) of the 
$\th$-expansion of the gravitino field-strength 
can be written schematically as
\begin{alignat}{3}
T^{(n)}&\sim\frac{\cO^{\frac{n}{2}}}{n!}\partial\Psi+U^{(n)}, &~~~~~n=2k~,\nn\\
T^{(n)}&\sim\frac{\cO^{\frac{n-1}{2}}}{n!}(\th R+\th\partial G)+U^{(n)}, 
&~~~~~n=2k+1~,\nn
\end{alignat}
where $U^{(n)}$  is a known expression nonlinear in the fields
and ${\cal O}$ is a (matrix) differential operator quadratic in $\theta$;  
schematically, 
$\cO\sim (\theta\C\th)\partial$.
We have denoted by $\Psi$, $R$, $G$, the gravitino, Riemann tensor and 
four-form field strength of 
eleven-dimensional supergravity, respectively. 
Since $U^{(n)}$ is nonlinear, the equations above can be 
iterated to any order in the number of fields. 
In other words: {\it perturbatively in the number of fields we can obtain 
expressions which are 
exact to all orders in} $\th$. 

In the following section we review the superspace 
formulation of eleven-dimensional supergravity. 
In section 3 we describe the gauge-fixing procedure. 
In section 4 we obtain recursion relations which are 
then iterated in section 5 to obtain the 
explicit expansion of the vielbein and 
three-form potential to 
order ${\cal O}(\th^5)$. The expansion 
in the number of fields is described in section 6. 
The linear coupling of the 
covariant supermembrane to the background fields, is given 
in section 7. Section 8 contains a discussion of 
future directions and possible applications of our results. 
Since this is a somewhat technical paper, 
for quick reference we have included an index of 
various definitions used, in appendix A. 
In appendix B we have included a note on our 
conventions concerning gamma matrices and spinor notation. 
Appendix C 
contains the vielbein and 
three-form potential expansions for
the simplified case of a purely geometric background.


\section{On-shell 11D supergravity in superspace}

This section is a summary of known results that can be found in 
or deduced from the literature. 
We have included it in order 
to establish notation and conventions, 
and to make the paper self-contained. 

Eleven-dimensional supergravity \cite{cjs} 
admits a superspace formulation \cite{cf, brh}. 
Let $A=(a,\alpha)$; $a=0\dots 10, ~\alpha=1\dots 32$, be a flat superspace index
and let $E^A=(E^a,E^\a)$ be the coframes of the $(11|32)$ supermanifold.
Moreover, let us introduce a connection one-form $\O_{A}{}^B$ with
Lorentzian structure group.
The supertorsion and supercurvature are given by
\begin{align}
T^A &=DE^A:=dE^A + E^B \O_B{}^A=\frac{1}{2} E^C E^B T_{BC}{}^A \nn\\
R_A{}^B &=d\O_A{}^B +\O_A{}^C\O_C{}^B=\frac{1}{2} E^D E^C R_{CD,A}{}^B
\label{2.1}
\end{align}
and obey the Bianchi identities 
\begin{align}
DT^A&=E^B R_B{}^A \nn\\
DR_B{}^A&=0~.
\la{parker}
\end{align}
Note that for a Lorentzian structure group the second  Bianchi identity 
follows from the first \cite{dragon}.
In a purely geometrical
definition in terms of the supertorsion, it was shown
in \cite{hw} that the equations of motion of 
11D supergravity follow from the constraint
\be
T_{\alpha\beta}{}^{a}= -i(\C^a)_{\alpha\beta}~.
\la{plato}
\end{equation}
In this formulation the physical fields of the theory, 
the graviton, the gravitino and the 
three-form potential, appear through their covariant field strengths. 
Namely, the curvature $R_{ab,c}{}^d$ is identified with the
top component of the supercurvature, the gravitino field-strength
$T_{ab}{}^\a$ 
is identified with the dimension three-halves component of the
supertorsion, while the four-form field strength $G_{abcd}$ 
appears in the dimension-one components 
of the supertorsion and supercurvature. 

More explicitly, 
the remaining 
nonzero components of the supertorsion and supercurvature 
of undeformed 11D supergravity are given by
\be
T_{a\beta}{}^{\c}= (\cT_a{}^{bcde})_\b{}^\c G_{bcde}
\label{2.4}
\end{equation}
and
\begin{align}
R_{\a\beta,ab} &=i(\cR_{ab}{}^{cdef})_{\a\b}G_{cdef}\nn\\
R_{\alpha b,cd}&=i(\cS_{bcd}{}^{ef})_{\a\b}T_{ef}{}^\b ,
\label{blirp}
\end{align}
where
\bead
\cT_a{}^{bcde}&:=-\frac{1}{36}\left(\d_a^{[b}\C^{cde]}
+\frac{1}{8}\C_a{}^{bcde}  \right)\nn\\
\cR_{ab}{}^{cdef}&:=\frac{1}{6}\left(\d_a^{[c}\d_b^{d}\C^{ef]}  
+\frac{1}{24}\C_{ab}{}^{cdef} \right)\nn\\
\cS_{bcd}{}^{ef}&:= \frac{1}{2}\left(
\C_b\d_c^{[e}\d_d^{f]}+\C_c\d_b^{[e}\d_d^{f]}-\C_d\d_b^{[e}\d_c^{f]} \right).
\label{ugm}
\end{alignat}
Note that the Lorentz condition implies
\be
R_{AB\a}{}^\beta=\frac{1}{4}R_{ABcd}(\C^{cd})_\a{}^\beta~.
\end{equation}
The action of the spinorial derivative on the physical
field strengths is given by
\begin{align}
\nabla_{\a}G_{abcd}&=6i(\C_{[ab}T_{cd]})_\a\nn\\
\nabla_{\a}T_{ab}{}^\beta&=\frac{1}{4}R_{ab,cd}(\C^{cd})_{\a}{}^\beta
-2\nabla_{[a}T_{b]\a}{}^\beta-2T_{[a|\a}{}^\epsilon 
T_{|b]\epsilon}{}^\beta\nn\\
\nabla_{\a}R_{ab,cd}
&=2\nabla_{[a|}R_{\a|b]cd}-T_{ab}{}^\epsilon R_{\epsilon \alpha cd}
+2T_{[a|\a}{}^\epsilon R_{\epsilon|b]cd} ~.
\label{derivatives}
\end{align}
The equations-of-motion that follow from 
(\ref{derivatives}) read
\begin{align}
\nabla_{[a}G_{bcde]}&=0\nn\\
\nabla^fG_{fabc}&=-\frac{1}{2(4!)^2}
\varepsilon_{abcd_1\dots d_8}G^{d_1\dots d_4}G^{d_5\dots d_8} \nn\\
(\C^aT_{ab})_\a&=0\nn\\
R_{ab}-\frac{1}{2}\eta_{ab}R&=-\frac{1}{12}\left(G_{adfg}G_b{}^{dfg}
-\frac{1}{8}\eta_{ab}G_{dfge}G^{dfge}  \right) ~.
\label{eqsofmo}
\end{align}
The equations above imply the existence of 
a closed superfour-form $G$ which can be
written locally in terms of a superpotential $C$,
\be
G=dC.
\label{2.5}
\end{equation}
The only nonvanishing components of $G$ are
\be
G_{abcd}, ~~~~ G_{\a\b ab}=-i(\C_{ab})_{\a\b}.
\label{2.6}
\end{equation}
In order to derive the equations of motion in component form, 
a little more work is required. 
Equations (\ref{2.5},\ref{2.6}) together with
\be
G_{MNPQ}=(-)^{m(c+n+b+p+a+q)}(-)^{n(b+p+a+q)}(-)^{p(a+q)}
E_Q{}^AE_P{}^BE_N{}^CE_M{}^DG_{DCBA}
\label{obione}
\end{equation}
imply
\be\boxed{
e_m{}^ae_n{}^be_p{}^ce_q{}^dG^{(0)}_{abcd}
= 4\partial_{[m}C^{(0)}_{npq]}
-6i(\Psi_{[m}\C_{np}\Psi_{q]})~,}
\label{four}
\end{equation}
where for any superfield $S$, 
\be
S^{(0)}:=S\vert_{\th=0}
\label{gotya}
\end{equation}
and we define
\bea
e_m{}^a&:=E^{(0)}_{~~m}{}^{a}\nn\\
\Psi_m{}^\a&:=E^{(0)}_{~~m}{}^{\a}\nn\\
\o_{mA}{}^{B}&:=\O^{(0)}_{mA}{}^{B}.
\label{koko}
\eea
It follows from definition (\ref{2.1}) that
\be
T^{(0)}_{mnk}=2(\partial_{[m}e_{n]}{}^a e_{ak}+\omega_{[mn]k}  ),
\label{2.7}
\end{equation}
where we have defined
\bead
T^{(0)}_{mnk}&:=T^{(0)}_{~~mn}{}^ae_{ak}\nn\\
\o_{mnk}&:=\o_{ma}{}^be_n{}^ae_{bk}.
\end{alignat}
On the other hand,
\be
T_{MN}{}^A=(-)^{m(n+c)}E_{N}{}^CE_{M}{}^BT_{BC}{}^A
\label{bla}
\end{equation}
together with (\ref{plato}) implies
\be
T^{(0)}_{~~mn}{}^a= i(\Psi_m\C^a\Psi_n),
\label{2.8}
\end{equation}
where we have suppressed spinor indices for simplicity.
Combining (\ref{2.7},\ref{2.8}) we arrive at
\be
\o_{nkm}=\oo_{nkm}+K_{nkm},
\label{2.9}
\end{equation}
where
\bead
\oo_{nkm}&:=
\partial_{[k}e_{m]}{}^ae_{an}
-\partial_{[m}e_{n]}{}^ae_{ak}-\partial_{[n}e_{k]}{}^ae_{am}\nn\\
K_{nkm}&:=\frac{i}{2}(\Psi_m\C_k\Psi_n+\Psi_n\C_m\Psi_k-\Psi_k\C_n\Psi_m)
\end{alignat}
and
\be
\C_m:=e_m{}^a\C_a.
\label{yuu}
\end{equation}
Note that the contorsion tensor $K$ satisfies
\be
K_{[mn]p}=\frac{1}{2}T^{(0)}_{mnp}.
\end{equation} 
With $\oo$ we 
can associate a covariant derivative $\cDo$ which obeys
\be
\cDo_n e_m{}^a := \partial_n  e_m{}^a 
- \Co{}^k_{nm} e_k{}^a +\oo_{nb}{}^a e_m{}^b=0,
\label{cona}
\end{equation}
where $\Co$ is the Levi-Civita connection.
Similarly, with $\o$ we can associate 
a covariant derivative $\cD$ which obeys
\be
\cD_n e_m{}^a := \partial_n  e_m{}^a 
- \C_{nm}^k e_k{}^a +\o_{nb}{}^a e_m{}^b=0,
\label{conb}
\end{equation}
where
\be
\C^p_{mn}=\Co{}^p_{mn}+K_{mn}{}^{p}.
\end{equation}
Proceeding as above, one can show that
\be\boxed{
e_m{}^ae_n{}^b~T^{(0)}_{~~ab}{}^\a
=\partial_{m}\Psi_n{}^\a+\o_{m\b}{}^\a\Psi_n{}^\b+
(\Psi_m\cT_n{}^{abcd})^\alpha G^{(0)}_{abcd}
-(m\leftrightarrow n)~,}
\label{tors}
\end{equation}
where 
\be
\cT_n{}^{abcd}:=e_n{}^f \cT_f{}^{abcd}.
\end{equation}
Finally, (\ref{2.1},\ref{2.8}) together with
\be
R_{MNA}{}^B=(-)^{m(n+d)}E_{N}{}^DE_{M}{}^CR_{CDA}{}^B
\label{lala}
\end{equation}
imply that
\be\boxed{
e_m{}^ae_n{}^be_k{}^ce_l{}^dR^{(0)}_{abcd}
=R(\o)_{mnkl}
+i(\Psi_m\cR_{kl}{}^{abcd} \Psi_n )G^{(0)}_{abcd}
-2i(\Psi_{[m}\cS_{n]kl}{}^{ab}T^{(0)}_{ab})~,}
\label{curv}
\end{equation}
where  
\bead
\cR_{kl}{}^{cdfg}&:=e_k{}^ae_l{}^b
\cR_{ab}{}^{cdfg}\nn\\
\cS_{nkl}{}^{cd}&:= e_n{}^fe_k{}^ae_l{}^bS_{fab}{}^{cd}.
\end{alignat}
We arrive at the equations of motion in component form
by taking (\ref{eqsofmo}) at $\theta=0$ 
and substituting (\ref{four},\ref{tors},\ref{curv}).


\section{Gauge-fixing}
The supervielbein contains an enormous amount of 
gauge freedom which can be fixed by using 
higher $\th$-levels of supersymmetry and 
Lorentz transformations. In this section we impose the 
most natural generalization, to 
all levels of the $\th$-expansion, of the Wess-Zumino gauge. 
This leads to the system of normal 
coordinates introduced in \cite{norcor} as a 
superspace generalization  
of the ordinary 
normal-coordinate expansion on Riemannian manifolds. 
As we will see below, significant 
simplifications occur when superspace quantities are expressed 
in these coordinates.

Let us first introduce some notation.
For any superfield $S_{\{ A\}}$ we define the coefficients in the 
$\theta$-expansion as follows:
\be
S_{\{ A\}}=\sum_{n=0}^{32}S_{\{ A\}}^{(n)},
\end{equation}
where
\be
S_{\{ A\}}^{(n)}:=\frac{1}{n!}\theta^{\m_n}\dots\theta^{\m_1}
S^{(n)}_{\m_1\dots\m_n, {\{ A\}}}
\label{gotyab}
\end{equation}
and ${\{ A\}}$ stands for all Lorentz indices.
In particular, the $\theta$-expansions 
of the vielbein and superconnection
read
\bead
E_M{}^A&=\sum_{n=0}^{32}E^{(n)}_{~~M}{}^A\nn\\
&=\sum_{n=0}^{32}\frac{1}{n!}\theta^{\m_n}\dots\theta^{\m_1}
E^{(n)}_{\m_1\dots\m_n, M}{}^A 
\end{alignat}
and 
\bead
\O_{MA}{}^B&=\sum_{n=0}^{32}\O^{(n)}_{MA}{}^B\nn\\
&=\sum_{n=0}^{32}\frac{1}{n!}\theta^{\m_n}\dots\theta^{\m_1}
\O^{(n)}_{\m_1\dots\m_n, MA}{}^B,
\end{alignat}
respectively. Under supersymmetry 
and Lorentz tranformations with parameters $\xi^A$ and 
$L_A{}^B$ respectively, the vielbein
and superconnection transform as \cite{wb}
\be
\d E_M{}^A=-\nabla_M\xi^A-\xi^B T_{BM}{}^A+E_M{}^B L_B{}^A
\label{vieltr}
\end{equation}
and
\be
\d \O_{MA}{}^B=-\xi^C R_{MCA}{}^B+\O_{MA}{}^C L_C{}^B 
-\O_{MC}{}^B L_A{}^C -\partial_ML_A{}^B,
\label{conntr}
\end{equation}
where 
\be
\nabla_M\xi^A:=\partial_M\xi^A+\O_{MB}{}^A\xi^B~.
\end{equation}
As can be seen from (\ref{vieltr}), we can use 
$\xi_{~~\mu,}^{(1)}{}^A$ to set
\bea
E^{(0)}_{~~\mu}{}^{a}&=0\nn\\
E^{(0)}_{~~\mu}{}^{\a}&=\d_\mu^{\a}.
\label{vielzero}
\eea
More generally, we can use 
$\xi_{\mu_1\dots\mu_{n+1},}^{(n+1)}{}^A$ 
to set 
to zero the totally antisymmetric parts
\be
E^{(n)}_{[\mu_1\dots\mu_{n},\mu ]}{}^{A}=0, ~~~~~1\leq n\leq 32.
\label{vielhigh}
\end{equation}
Similarly, from (\ref{conntr}) we see that 
we can use $L^{(1)}_{\mu, A}{}^B$ to set 
\be
\O^{(0)}_{~~\mu A}{}^{B}=0.
\label{connzero}
\end{equation}
The higher $\theta$-levels of the Lorentz 
transformation, $L^{(n+1)}_{\mu_1\dots\mu_{n+1}, A}{}^B$, 
can be used to set
\be
\O^{(n)}_{[\mu_1\dots\mu_{n},\mu ] A}{}^{B}=0, ~~~~~1\leq n\leq 32.
\label{connhigh}
\end{equation}
The gauge-fixing conditions (\ref{vielzero}-\ref{connhigh}) imply
\bead
\theta^\mu E_\mu{}^a&=0\nn\\
\theta^\mu E_\mu{}^\a&=\theta^\a
\label{1}
\end{alignat}
and
\be
\theta^\mu \O_{\mu A}{}^B=0,
\label{2}
\end{equation}
where we have defined
\be
\theta^\a:=\theta^\mu \d_\mu^\a.
\label{kalos}
\end{equation}
Expressions (\ref{1}),(\ref{2}) are  
identical to (A3),(A4) of \cite{norcor}.   

The inverse $E_A{}^M$ of the vielbein satisfies
\be
E_A{}^M E_M{}^B=\d_A^B.
\end{equation}
At zeroth-order in the $\theta$-expansion we have
\bead
E^{(0)}_{~~\a}{}^m&=0\nn\\
E^{(0)}_{~~\a}{}^\m&=\d_\a^\m\nn\\
E^{(0)}_{~~a}{}^\m&=-\Psi_a{}^\m:=-e_a{}^m\Psi_m{}^\a\d_\a^\m\nn\\
E^{(0)}_{~~a}{}^m&=e_a{}^m.
\label{3}
\end{alignat}
At order $n\geq 1$ we get
\bead
E^{(n)}_{~~A}{}^m&=-\sum_{r=0}^{n-1}E_{~~A}^{(r)}{}^M 
E^{(n-r)}_{~~~~M}{}^ae_a{}^m \nn\\
E^{(n)}_{~~A}{}^\mu&=\sum_{r=0}^{n-1}\left( 
-E_{~~A}^{(r)}{}^M E^{(n-r)}_{~~~~M}{}^\a\d_\a^\m 
+ E_{~~A}^{(r)}{}^M E^{(n-r)}_{~~~~M}{}^a\Psi_a{}^\m\right).
\label{4}
\end{alignat}
Using (\ref{3},\ref{4},\ref{1}) 
it is straightforward to prove by induction that
\bead
\theta^\alpha E_\a{}^m&=0\nn\\
\theta^\alpha E_\a{}^\m&=\theta^\m.
\label{5}
\end{alignat}
Moreover, from (\ref{5},\ref{2}) it follows that 
\be
\theta^\alpha \O_{\alpha A}{}^B=0.
\label{6}
\end{equation}
We also have
that for any superfield $S$ (suppressing
any Lorentz indices)
\bead
\theta^\a\nabla_\alpha S&=\theta^\alpha E_\a{}^M(\partial_M+\O_M )S \nn\\
&=\theta^\m\partial_\m S
\label{coop}
\end{alignat}
and therefore
\be
S^{(n)}=\frac{(n-r)!}{n!}~\theta^{\a_r}\dots\theta^{\a_1}
\left(\nabla_{\a_1}\dots\nabla_{\a_r}S\right)^{(n-r)}.
\label{7}
\end{equation}
Equation (\ref{coop}) is the same as equation (1) of 
\cite{norcor}.


\section{Recursion}

\subsection{$G_{abcd}$,$T_{ab}{}^\a$ and $R_{abcd}$}

Using the results in the previous sections 
we can now 
obtain recursion relations that relate different 
levels in the $\th$-expansions of the various superfields. 
First let us first derive 
the recursion relations for $G,T,R$. These are 
readily obtained from (\ref{7}),(\ref{derivatives}):
\bead
G_{abcd}^{(n)}&=\frac{6i}{n}(\theta\C_{[ab}T^{(n-1)}_{cd]})\nn\\
T^{(n)}_{~~ab}{}^\a&=\frac{1}{4n}(\theta\C^{cd})^\alpha R^{(n-1)}_{abcd}
+\frac{2}{n}(\theta\cT_{[a}{}^{cdef})^\alpha (\nabla_{b]}G_{cdef})^{(n-1)} \nn\\
&~~~~~~-\frac{2}{n}(\theta\cT_{[a}{}^{cdef}\cT_{b]}{}^{c'd'e'f'})^\a
(G_{cdef}G_{c'd'e'f'} )^{(n-1)}\nn\\
R^{(n)}_{abcd}&=
-\frac{2i}{n}(\theta\cS_{[a|cd}{}^{ef})_\a(\nabla_{|b]}T_{ef}{}^\a)^{(n-1)}
-\frac{i}{n}(\theta\cR_{cd}{}^{efgh})_\a(T_{ab}{}^\alpha G_{efgh})^{(n-1)}\nn\\
&~~~~~~+\frac{2i}{n}(\theta\cT_{[a}{}^{efgh} \cS_{b]cd}{}^{e'f'})_\a
(T_{e'f'}{}^\alpha G_{efgh})^{(n-1)}
~.
\label{grtrec}
\end{alignat}
For later use let us also note that
\be
T^{(n)}_{~~ab}{}^\a~ =\frac{i}{n(n-1)}\{ (\cM_{[a|}{}^{ef})^\a{}_\b
(\nabla_{|b]}T_{ef}{}^\b )^{(n-2)}
+(\cN_{ab}{}^{c_1\dots c_6})^\a{}_\b(G_{c_1\dots c_4}
T_{c_5c_6}{}^\b )^{(n-2)}
\},
\label{roufa1}
\end{equation}
where
\bead
(\cM_{a}{}^{ef})^\a{}_\b&:= 
-\frac{1}{2}(\theta\C^{bc})^\a(\theta\cS_{abc}{}^{ef})_\b
+12(\theta\cT_a{}^{bcef})^\a(\theta\C_{bc})_\b\nn\\
(\cN_{ab}{}^{c_1\dots c_6})^\a{}_\b&:= 
-\frac{1}{4}(\theta\C^{ef})^\alpha (\theta\cR_{ef}{}^{c_1\dots c_4})_\b
\d^{c_5}_{[a}\d^{c_6}_{b]}
+\frac{1}{2}(\theta\C^{ef})^\a(\theta\cT_{[a}{}^{c_1\dots c_4}
\cS_{b]ef}{}^{c_5c_6})_\b\nn\\
&~~~~+8(\theta\cT_{[a}{}^{ec_1c_2c_3})^\a
(\theta\cS_{b]e}{}^{c_4c_5c_6})_\b
+12(\theta\cT_{[a}{}^{efc_5c_6})^\a
(\theta\cT_{b]}{}^{c_1\dots c_4}\C_{ef})_\b\nn\\
&~~~~-12(\theta\cT_{[a}{}^{efc_5c_6}\cT_{b]}{}^{c_1\dots c_4})^\alpha 
(\theta\C_{ef})_\b
-12(\theta\cT_{[a}{}^{c_1\dots c_4}\cT_{b]}{}^{efc_5c_6})^\alpha 
(\theta\C_{ef})_\b~.
\label{roufa2}
\end{alignat}
This formula is most easily derived as follows: 
act with $\nabla_{[\a_1}\nabla_{a_2]}$ on $T_{ab}{}^\a$, 
taking (\ref{derivatives}) into account; substitute the 
result in (\ref{7}) for $S\rightarrow T_{ab}{}^\a$, 
$r\rightarrow 2$.


\subsection{$\O_M$, $E_M{}^A$ and $E_A{}^M$}

We now turn to the derivation of recursion 
relations for the components of the connection and the vielbein.   
The first line of definition (\ref{2.1}) implies
\be
2\partial_{(\mu}E_{\nu)}{}^a=
T_{\mu\nu}{}^a-2\O_{(\mu|e}{}^aE_{|\nu)}{}^e.
\end{equation}
Multiplying both sides by $\theta^\mu$, 
taking (\ref{1},\ref{2}) into account, gives
\be
E^{(n+1)}_{~~~~\mu}{}^a 
=\frac{1}{n+2}\theta^\n T_{\n\m}{}^a,~~~~~n\geq 0~.
\end{equation}
Moreover, using (\ref{plato}), (\ref{bla}) we arrive at
\be\boxed{
E^{(n+1)}_{~~~~\mu}{}^a 
=-\frac{i}{n+2}E^{(n)}_{~~\mu}{}^\a(\C^a\theta)_\a,~~~~~n\geq 0~.}
\label{rec1}
\end{equation}
Similarly, we can show 
that 
\bead\boxed{
E^{(n+1)}_{~~~~m}{}^a = 
-\frac{i}{n+1}E^{(n)}_{~~m}{}^\a(\C^a\theta)_\a,~~~~~n\geq 0~.}
\label{rec2}
\end{alignat}
The second line of definition (\ref{2.1}) implies
\be
2\partial_{(\mu}\O_{\nu)a}{}^b=
R_{\mu\nu a}{}^b+2\O_{(\mu|a}{}^c\O_{|\nu)c}{}^b.
\end{equation}
Multiplying both sides by $\theta^\mu$
gives
\bead
\O^{(n+1)}_{~~~~\mu ab} &= \frac{1}{n+2}~\theta^\nu 
R^{(n)}_{~~~\nu\mu ab}\nn\\
&=\frac{i}{n+2}\sum_{r=0}^{n}\{ E^{(r)}_{~~\m}{}^\alpha 
(\cR_{ab}{}^{cdef}\theta)_\a
G^{(n-r)}_{cdef}\nn\\
&~~~~~~~~~~~~~~~~~~~~~+E^{(r)}_{~~\m}{}^e(\theta \cS_{eab}{}^{cd}
T^{(n-r)}_{cd})\}
, ~~~~~n\geq 0~.
\end{alignat}
The first equality is shown by taking (\ref{2}) into account.
The second one is a consequence of (\ref{lala},\ref{1},\ref{blirp}). 
Furthermore, using the recursion relations (\ref{grtrec}),(\ref{rec1}) 
yields
\bead
\O^{(n+1)}_{~~~~\mu ab} &= 
\frac{i}{n+2}E^{(n)}_{~~\m}{}^\alpha 
(\theta\cR_{ab}{}^{cdef})_\alpha 
G^{(0)}_{cdef} \nn\\
&+\frac{1}{n+2}\sum_{r=0}^{n-1}E^{(r)}_{~~\m}{}^\alpha 
\big( \frac{1}{n-r}C_{1ab}{}^{ef}+ \frac{1}{r+2}C_{2ab}{}^{ef}
\big)_{\a\b}T^{(n-r-1)}_{~~~~~~~ef}{}^\b , ~~~~~n\geq 0~,
\label{rec3}
\end{alignat}
where
\bead
(C_{1ab}{}^{ef})_{\a\b }&
:=-6(\theta\cR_{ab}{}^{cdef})_\a(\theta\C_{cd})_\b
\nn\\
(C_{2 ab}{}^{ef})_{\a\b }&:=(\theta\C^g)_\a(\theta \cS_{gab}{}^{ef})_\b~.
\label{cdefs}
\end{alignat}
Note that $\theta^\mu\O_\mu=0$, 
as it should. This follows from (\ref{1}) and the symmetry of 
$\cR$.

Similarly one can show that
\bead
\O^{(n+1)}_{~~~~m ab} &= 
\frac{i}{n+1}
(\theta\cS_{mab}{}^{ef}T^{(n)}_{~~ef})
+\frac{i}{n+1}E^{(n)}_{~~m}{}^\alpha 
(\theta\cR_{ab}{}^{cdef})_\a
G^{(0)}_{cdef} \nn\\
&+\frac{1}{n+1}\sum_{r=0}^{n-1}E^{(r)}_{~~m}{}^\alpha 
\big( \frac{1}{n-r}C_{1ab}{}^{ef}+ \frac{1}{r+1}C_{2ab}{}^{ef}
\big)_{\a\b}T^{(n-r-1)}_{~~~~~~~ef}{}^\b , ~~~~~n\geq 0~.
\label{omrec}
\end{alignat}
We are now ready to obtain 
recursion relations for the remaining components of the vielbein. 
The first line of definition (\ref{2.1}) implies
\be
2\partial_{(\mu}E_{\nu)}{}^\a=
T_{\mu\nu}{}^\a-2\O_{(\mu|\b}{}^\alpha E_{\nu)}{}^\b.
\end{equation}
Multiplying both sides by $\theta^\mu$, 
taking (\ref{1},\ref{2}) into account, gives
\bead
E^{(n+1)}_{~~~~~\mu}{}^\alpha &= \frac{1}{n+2}
\left(\theta^\nu T^{(n)}_{~~~\nu\mu}{}^\alpha 
-\theta^\b\O^{(n)}_{~~\mu\b}{}^\alpha  \right)\nn\\
&=\frac{1}{n+2}~
\theta^\b\left( \sum_{r=0}^nE^{(r)}_{~~\n}{}^aT^{(n-r)}_{~~~a\b}{}^\alpha 
-\O^{(n)}_{~~\mu\b}{}^\alpha  \right)
 ,~~~~~n\geq 0~.
\label{prip}
\end{alignat}
The second equality follows from (\ref{bla}).
{}For $n=0$ in particular we have
\be\boxed{
E^{(1)}_{~~\mu}{}^\alpha =0~,}
\label{rec41}
\end{equation}
as can be seen from (\ref{vielzero}),(\ref{connzero}).
For $n\geq 1$, we can further reduce (\ref{prip}) by using 
(\ref{rec3}),(\ref{rec1}),(\ref{2.4})
\begin{center}
\fbox{\parbox{15.6cm}{
%
\bead
E^{(n+1)}_{~~~~~\mu}{}^\alpha &= 
\frac{i}{(n+1)(n+2)}
E^{(n-1)}_{~~~~~\mu}{}^\b (D_1^{cdef})_\b{}^\alpha G^{(0)}_{cdef}\nn\\
&+\frac{1}{(n+1)(n+2)}
\sum_{r=0}^{n-2}E^{(r)}_{~~\mu}{}^\b 
\big(
\frac{1}{n-r-1}F_1^{ef}
+\frac{1}{r+2}F_2^{ef}\nn\\
&~~~~~~~~~~~~~~~~~~~~~~~~~~~~~~~
~~~~~~~~~~~~+\frac{n+1}{(n-r-1)(r+2)}F_3^{ef}
\big)^\a{}_{\b\c}T^{(n-r-2)}_{~~~~~~~~ef}{}^\c
,~~~~~n\geq 1~,\nn
\end{alignat}
%
}}
\end{center}
\be\label{rec42}\end{equation}
where
\bead
(D_1^{cdef})_\b{}^\a&:=
\frac{1}{4}(\theta\cR_{ab}{}^{cdef})_\b(\theta\C^{ab})^\a
+(\theta\C^{a})_\b(\theta\cT_{a}{}^{cdef})^\a\nn\\
(F_1^{ef})^\a{}_{\b\c}&:=\frac{3}{2}(\theta\C^{ab})^\a
(\theta\cR_{ab}{}^{cdef})_\b(\theta\C_{cd})_\c\nn\\
(F_2^{ef})^\a{}_{\b\c}&:=-\frac{1}{4}(\theta\C^{ab})^\a
(\theta\C^g)_\b(\theta \cS_{gab}{}^{ef})_\c \nn\\
(F_3^{ef})^\a{}_{\b\c}&:=
6(\theta\cT_a{}^{bcef})^\a(\theta\C^a)_\b(\theta\C_{bc})_\c  ~.
\label{dfdefs}
\end{alignat}
Note that for $n>0$ it follows that 
$\theta^\mu E^{(n)}_{~~\mu}{}^\a=0$, as it should.

We proceed similarly to arrive at the following recursion relations:
\be\boxed{
E^{(1)}_{~~m}{}^\alpha = \frac{1}{4}(\theta\C^{ab})^\a\o_{mab}
-(\theta\cT_m{}^{cdef})^\alpha G^{(0)}_{cdef}}
\label{rec51}
\end{equation}
and
\begin{center}
\fbox{\parbox{15.6cm}{
%
\bead
E^{(n+1)}_{~~~~m}{}^\alpha &= 
\frac{i}{n(n+1)}E^{(n-1)}_{~~~~~m}{}^\b 
(D_1^{cdef})_\b{}^\alpha G^{(0)}_{cdef}
+\frac{i}{n(n+1)}T^{(n-1)}_{~~~~~ef}{}^\b (D_{2m}{}^{ef})_\b{}^\alpha \nn\\
&+\frac{1}{n(n+1)}
\sum_{r=0}^{n-2}E^{(r)}_{~~m}{}^\b 
\big(
\frac{1}{n-r-1}F_1^{ef}
+\frac{1}{r+1}F_2^{ef}\nn\\
&~~~~~~~~~~~~~~~~~~~~~~~~~~~~~~~
~~~~~~~~~~~~+\frac{n}{(n-r-1)(r+1)}F_3^{ef}
\big)^\a{}_{\b\c} T^{(n-r-2)}_{~~~~~~~~ef}{}^\c
,~~~~~n\geq 1 ~,\nn
\label{rec52}
\end{alignat}
%
}}
\end{center}
\be\label{rec52}\end{equation}
where
\be
(D_{2a}{}^{bc})_\b{}^\a:=
-\frac{1}{4}(\theta \cS_{aef}{}^{bc})_\b(\theta\C^{ef})^\a
+6(\theta\C_{ef})_\b(\theta \cT_{a}{}^{bcef})^\alpha ~.
\label{ddef}
\end{equation}
Finally, note that if $E_M{}^A$ is known 
to order $\theta^n$ and $E_A{}^M$ is known 
to order $\theta^{n-1}$, then equations (\ref{4}) can be used to
derive  $E_A{}^M$ to order $\theta^n$.

\subsection{$C_{MNP}$}

Let us turn to the recursion
relations for the components of the three-form superpotential.
The results of this section will be relevant to 
the derivation of the membrane action 
in section 7.

By definition, the $C$-field satisfies
\be
4\partial_{[M}C_{NPQ\}}=G_{MNPQ}.
\label{opiopi}
\end{equation}
Up to a gauge choice, 
the following is a solution of
(\ref{opiopi}) at each order 
in the $\theta$ expansion
\bead
C^{(0)}_{\m\n\s}=
C^{(0)}_{\m\n s}&=
C^{(0)}_{\s mn}=0~,\nn\\
4\partial_{[m}C^{(0)}_{npq]}&=G^{(0)}_{mnpq}
\label{cexpa}
\end{alignat}
and
\bead
C^{(n+1)}_{\m\n\s}&=\frac{1}{n+4}~\theta^\l G^{(n)}_{\l\m\n\s}\nn\\
C^{(n+1)}_{\m\n s}&=\frac{1}{n+3}~\theta^\l G^{(n)}_{\l\m\n s}\nn\\
C^{(n+1)}_{\s mn}&=\frac{1}{n+2}~\theta^\l G^{(n)}_{\l\s mn}\nn\\
C^{(n+1)}_{mnp}&=\frac{1}{n+1}~\theta^\l G^{(n)}_{\l mnp}
~, ~~~~~n\geq 0~.
\label{cexp}
\end{alignat}
Together with (\ref{obione}), equations 
(\ref{cexpa}), (\ref{cexp})
are enough to determine 
the $\theta$ expansion of the $C$-field.

The right-hand sides of
the formulae above can be further reduced by taking 
(\ref{obione}),(\ref{2.6}),(\ref{1}) 
into account. 
Explicitly we find
\bead
\theta^\l G_{\l\m\n\s}&=-3iE_{(\m}{}^a E_\n{}^b
E_{\s)}{}^\d(\C_{ab}\theta)_\d \nn\\
\theta^\l G_{\l\m\n s}&=-i E_\m{}^a E_\n{}^b E_s{}^\c
(\C_{ab}\theta)_\c-2i E_s{}^a E_{(\m}{}^b E_{\n)}{}^\c(\C_{ab}\theta)_\c\nn\\
\theta^\l G_{\l\s mn}&=-i E_m{}^a E_n{}^b E_\s{}^\d(\C_{ab}\theta)_\d
-2i E_\s{}^a E_{[m}{}^b E_{n]}{}^\c(\C_{ab}\theta)_\c\nn\\
\theta^\l G_{\l mnp}&=-3i E_{[m}{}^a E_n{}^b E_{p]}{}^\c(\C_{ab}\theta)_\c
~.
\label{cexpr}
\end{alignat}
%

\subsection{Maximally-supersymmetric superspaces}

The recursion relations of the previous section can be 
solved iteratively, 
order-by-order in an expansion in powers of $\theta$. We carry out
 this procedure in the 
next section, up to order $\cO(\theta^5)$. As mentioned in 
the introduction, 
around flat space a `dual' perturbative expansion 
is also possible: exact in $\theta$ but 
perturbative in powers of the background fields. This is explained 
in detail in section 6. However, it has been known for some time 
that in the special 
case of maximally-supersymmetric 
bosonic backgrounds of the type 
$AdS_d\times S^{D-d}$ one can obtain exact, closed expressions
for the supervielbien components \cite{krr, dffftt, wpps, claus, mt}. 
We now rederive this result using the methods of the present paper. 

We first note that in a bosonic background $T^{(0)}_{~~ab}{}^{\a}$ vanishes.  
Moreover, using (\ref{grtrec}),(\ref{conb}) 
it can be seen that the first $\theta$ level
of the gravitino field strength is given by
\begin{align}
T^{(1)}_{~~ab}{}^\a&=
e_a{}^me_b{}^n\{
\frac{1}{4}(\theta\C^{pq})^\alpha R(\o)_{mnpq}
+2(\theta\cT_{[m}{}^{pqrs})^\alpha (\cD_{n]}G_{pqrs})\nn\\
&~~~~~~~~~~~~~~~~-2(\theta\cT_{[m}{}^{pqrs}\cT_{n]}{}^{p'q'r's'})^\a
G_{pqrs} G_{p'q'r's'}
\}\nn\\
&=e_a{}^me_b{}^n\theta^\beta(\cR_{mn}^{Tr})_{\beta}{}^\a  ~,
\label{fgh}
\end{align}
where $G_{pqrs}:=4\partial_{[p}C^{(0)}_{qrs]} $ and 
$(\cR_{mn})^\a{}_\beta :=\big( [\mathbb{D}_m,\mathbb{D}_n]\big)^\a{}_\beta$ 
is the curvature of the supercovariant derivative
\be
(\mathbb{D}_m)^\a{}_\beta:=(\cD_m)^\a{}_{\beta}
-(\cT^{Tr}_{~~m}{}^{pqrs})^\a{}_\beta G_{pqrs}~.
\end{equation}
Note that 
in a bosonic background $\o=\oo$ and $\cD=\cDo$. Killing spinors are 
parallel with respect to $\mathbb{D}$. It follows that 
Killing spinors are  
eigenverctors of the supercurvature ($\cR_{mn}$) 
with zero eigenvalue and therefore 
in a maximally-supersymmetric space
\be
(\cR_{mn})^\a{}_\beta =0~,
\label{maxi}
\end{equation}
so that $T^{(1)}_{~~ab}{}^\a$ vanishes by (\ref{fgh}). 
All possible 
solutions to (\ref{maxi}) were classified in \cite{fp} up to 
local isometries. These are: $\mathbb{R}^{1,10}$, $AdS_4\times S^7$, 
$AdS_7\times S^4$, and the $Hpp$-wave. 

Since $T^{(n)}_{~~ab}{}^\alpha$ is zero for $n=0,1$, 
it now follows by induction from (\ref{roufa1}) 
that $T_{ab}{}^\alpha$ vanishes identically. 
Hence in the special case of maximally-supersymmetric spaces 
the gravitino field-strength drops out of the recursion 
relations (\ref{rec42}), (\ref{rec52}), 
which can now be solved straightforwardly. 
The result is
\begin{align}
E_\mu{}^\a&=\delta_\mu{}^\beta [\cP^{-1/2}sinh\sqrt{\cP}]_\beta{}^\a\nn\\
E_m{}^\a&=E^{(1)}_{~~m}{}^\beta [\cP^{-1/2}sinh\sqrt{\cP}]_\beta{}^\a~,
\label{maxa}
\end{align}
where 
\be
[\cP]_\a{}^\beta:=i(D_1^{mnpq})_\a{}^\beta G_{mnpq}~,
\end{equation}
and the functions of $\cP$ above are defined formally by their Taylor 
expansions around zero. The expressions for $D_1^{mnpq}
:=e_a{}^me_b{}^ne_c{}^pe_d{}^qD_1^{abcd}$ and $E^{(1)}_{~~m}{}^\a$ 
were given in (\ref{dfdefs}),(\ref{rec51}) respectively. 
The remaining components of the supervielbein follow similarly from  
(\ref{rec1}),(\ref{rec2}):
\begin{align}
E_\mu{}^a&=2i\delta_\mu{}^\beta 
[\cP^{-1}sinh^2\frac{\sqrt{\cP}}{2}]_\beta{}^\a
(\C^a\theta)_\a\nn\\
E_m{}^a&=e_m{}^a+2i E^{(1)}_{~~m}{}^\beta 
[\cP^{-1}sinh^2\frac{\sqrt{\cP}}{2}]_\beta{}^\a
(\C^a\theta)_\a~.
\label{maxb}
\end{align}
It can be seen that equations (\ref{maxa}),(\ref{maxb}) above are the same as 
(up to conventions) equations (22),(23) of \cite{krr}, 
or equation (3.9) of \cite{wpps}\footnote{
The matrix $\cM^2$ in those references is proportional 
to the matrix $\cP$ of the present paper.}. 
The method presented here like 
\cite{claus}  
does not rely on coset-superspace techniques. As is clear from the 
derivation, formulae (\ref{maxa}),
(\ref{maxb}) 
are  valid for all eleven-dimensional  
maximally-supersymmetric bosonic backgrounds.


\section{Expansion up to $\cO(\theta^5)$}

We now give the  explicit expressions for the first few terms in the
$\th$-expansion of the vielbein and the $C$-field.
Explicit (partial) results to order $\theta^2$ 
have appeared in \cite{plef} (see also \cite{kiva}). 
Our results are in complete agreement with those of \cite{plef} 
(cf. equations (4.15),(4.16) of that reference) to the order 
they have computed 
\footnote{In order to translate the expressions of \cite{plef} 
to our conventions, one needs to make the following substitutions: 
$\th\rightarrow \frac{1}{\sqrt{2}}\th$; $\bar{\th}\rightarrow 
\frac{i}{\sqrt{2}}\th^{Tr}C$
; $\Psi_m\rightarrow \frac{1}{\sqrt{2}}\Psi_m$ and   
$E_\m{}^a\rightarrow {\sqrt{2}}E_\m{}^a$; 
$E_m{}^\a\rightarrow \frac{1}{\sqrt{2}}
 E_m{}^\a$; $B_{mnp}\rightarrow -C_{mnp}$; 
$B_{mn\s}\rightarrow -{\sqrt{2}} C_{mn\s}$;  
$B_{m\n\s}\rightarrow -{2} C_{m\n\s}$; 
$B_{\m\n\s}\rightarrow -{2\sqrt{2}}
C_{\m\n\s}$; $\hat{F}_{mnpq}\rightarrow -G^{(0)}_{mnpq}$; 
$T_s{}^{mnpq}\rightarrow (\cT^{Tr})_s{}^{mnpq}$.}.

The authors of \cite{gris} have computed the vielbein expansion
to ${\cal O}(\theta^3)$, but have omitted some 
tedious (albeit straightforward) steps that would 
facilitate the comparison of their expressions 
to the explicit formulae in this section. After 
some algebra, formulae (5.7)-(5.10) of \cite{gris} can be seen to  
agree with our results (to the order they have computed), apart from 
signs that may be attributed to different conventions 
\footnote{Our $\th$ is equivalent to $y$ of \cite{gris}. 
Note that there seem to be the following typos in \cite{gris}: 
${\cal O}(y^5)$ in (5.8) should be replaced by ${\cal O}(y^4)$, 
${\cal O}(y^4)$ in (5.10) should be replaced by ${\cal O}(y^3)$,
the factor of $1/2$ in front of the second term on the 
right-hand side of (5.10) should 
be replaced by $1/6$, the $i$ in front of the second term on the 
right-hand side of (5.9) should 
be deleted.}.

In the following we will need the
explicit expressions for $T^{(1)}$, $T^{(2)}$.
These can be read off from (\ref{grtrec}),(\ref{roufa1}):
\bead
T^{(1)}_{~~ab}{}^\a&=\frac{1}{4}(\theta\C^{cd})^\alpha R^{(0)}_{abcd}
+2(\theta\cT_{[a}{}^{cdef})^\alpha (\nabla_{b]}G_{cdef})^{(0)}\nn\\
&~~~~~~~~~~~~~~~~-2(\theta\cT_{[a}{}^{cdef}\cT_{b]}{}^{c'd'e'f'})^\a
G_{cdef}^{(0)}G_{c'd'e'f'}^{(0)}~,
\label{deft1}
\end{alignat}
where
\bead
(\nabla_{b}G_{cdef})^{(0)}&=
e_b{}^m\cD_m G_{cdef}^{(0)}
+E^{(0)}_{~~b}{}^\mu \partial_\mu G_{cdef}^{(1)}
\nn\\
&=e_b{}^m\cD_m G_{cdef}^{(0)}
-6i(\Psi_b\C_{[cd}T^{(0)}_{ef]} )
\end{alignat}
and
\be
T^{(2)}_{~~ab}{}^\a~ =\frac{i}{2}(\cM_{[a|}{}^{ef})^\a{}_\b
(\nabla_{|b]}T_{ef}{}^\b )^{(0)}
+\frac{i}{2}(\cN_{ab}{}^{c_1\dots c_6})^\a{}_\b G^{(0)}_{c_1\dots c_4}
T^{(0)}_{~~c_5c_6}{}^\b ~,
\label{deft2}
\end{equation}
where
\bead
(\nabla_{b}T_{ef}{}^\b)^{(0)}&=
e_b{}^m\cD_m T^{(0)}_{~~ef}{}^\b
+E^{(0)}_{~~b}{}^\mu \partial_\mu T^{(1)}_{~~ef}{}^\b
\nn\\
&=e_b{}^m\cD_m T^{(0)}_{~~ef}{}^\b
-\frac{1}{4}(\Psi_b\C^{cd})^\b R^{(0)}_{efcd}\nn\\
&-2(\Psi_b\cT_{[e}{}^{cdgh})^\b e_{f]}{}^m\cD_m G^{(0)}_{cdgh}
+12i(\Psi_b\cT_{[e}{}^{cdgh})^\b (\Psi_{f]}\C_{cd}T^{(0)}_{gh} )\nn\\
&+2(\Psi_b\cT_{[e}{}^{cdgh}T_{f]}{}^{c'd'g'h'})^\b G^{(0)}_{cdgh}
G^{(0)}_{c'd'g'h'}
~.
\end{alignat}
\subsection*{Vielbein expansions}

Using formulae (\ref{rec1}),(\ref{rec2}),(\ref{rec41}),(\ref{rec42}),
(\ref{rec51}),(\ref{rec52}) 
we find for the first few terms
\bead
E^{(0)}_{~~\m}{}^a&=0\nn\\
E^{(1)}_{~~\m}{}^a&=-\frac{i}{2}(\C^a\theta)_\m\nn\\
E^{(2)}_{~~\m}{}^a&=0\nn\\
E^{(3)}_{~~\m}{}^a&=\frac{1}{24}(D_1^{cdef}\C^a\theta)_\m 
G^{(0)}_{cdef}\nn\\
E^{(4)}_{~~\m}{}^a&=\frac{i}{120}
\Big(2F_1^{ef}+F_2^{ef}+3F_3^{ef} 
\Big)^\a{}_{\m\b} (\C^a\theta)_\alpha T_{ef}^{(0)}{}^\b \nn\\
E^{(5)}_{~~\m}{}^a&= \frac{i}{720}(D_1^{cdef}
D_1^{c'd'e'f'}\C^a\theta)_{\m}{} G^{(0)}_{cdef}G^{(0)}_{c'd'e'f'}
+\frac{i}{240}\Big(F_1^{ef}+F_2^{ef}+2F_3^{ef} 
\Big)^\a{}_{\m\b}(\C^a\theta)_\alpha T_{ef}^{(1)}{}^\b~,
\end{alignat}
\bead
E^{(0)}_{~~m}{}^a&=e_m{}^a\nn\\
E^{(1)}_{~~m}{}^a&=-i(\Psi_m\C^a\theta)\nn\\
E^{(2)}_{~~m}{}^a&=-\frac{i}{8}(\theta\C^{aef}\theta)\o_{mef}
+\frac{i}{2}(\theta\cT_m{}^{cdef}\C^{a}\theta) 
G^{(0)}_{cdef}\nn\\
E^{(3)}_{~~m}{}^a&=\frac{1}{6}(\Psi_m D_1^{cdef}\C^a\theta) G^{(0)}_{cdef}
+\frac{1}{6}(T_{ef}^{(0)}D_{2m}{}^{ef}\C^a\theta)\nn\\
E^{(4)}_{~~m}{}^a&=\frac{1}{96}(\theta\C^{gh} D_1^{cdef}
\C^a\theta) \o_{mgh} G^{(0)}_{cdef}
-\frac{1}{24}(\theta \cT_{m}{}^{cdef}D_{1}^{c'd'e'f'}\C^a\theta) 
G^{(0)}_{cdef}G^{(0)}_{c'd'e'f'}  \nn\\
&+\frac{1}{24}(T_{ef}^{(1)}D_{2m}{}^{ef}\C^a\theta)
+\frac{i}{24}\Big(F_1^{ef}+F_2^{ef}+2F_3^{ef} 
\Big)^\a{}_{\b\c}(\C^a\theta)_\a\Psi_m{}^\b T_{ef}^{(0)}{}^\c\nn\\
E^{(5)}_{~~m}{}^a&=\frac{i}{120}(\Psi_m D_1^{cdef}
D_1^{c'd'e'f'}\C^a\theta) G^{(0)}_{cdef}G^{(0)}_{c'd'e'f'}
+\frac{i}{120}(T_{ef}^{(0)}D_{2m}{}^{ef} D_1^{cdgh}\C^a\theta)
G^{(0)}_{cdgh}\nn\\
&+\frac{i}{120}\Big(F_1^{ef}+2F_2^{ef}+3F_3^{ef} 
\Big)^\a{}_{\b\c}(\C^a\theta)_\a\Psi_m{}^\b T_{ef}^{(1)}{}^\c
+\frac{1}{60}(T^{(2)}_{ef} D_{2m}{}^{ef}\C^a\theta)\nn\\
&+\frac{i}{120}
\Big(2F_1^{ef}+F_2^{ef}+3F_3^{ef} 
\Big)^\a{}_{\b\c}
(\C^a\theta)_\a
(
\frac{1}{4}\theta\C^{gh}\o_{mgh}
-\theta\cT_m{}^{cdgh}
G^{(0)}_{cdgh} )^\b 
T_{ef}^{(0)}{}^\c
\end{alignat}
and
\bead
E^{(0)}_{~~\m}{}^\a&=\d_\m{}^\a\nn\\
E^{(1)}_{~~\m}{}^\a&=0\nn\\
E^{(2)}_{~~\m}{}^\a&=\frac{i}{6}(D_{1}^{cdef})_\m{}^\alpha G^{(0)}_{cdef}\nn\\
E^{(3)}_{~~\m}{}^\a&=\frac{1}{24}\Big(2F_1^{ef}+F_2^{ef}+3F_3^{ef} 
\Big)^\a{}_{\m\b}T_{ef}^{(0)}{}^\b \nn\\
E^{(4)}_{~~\m}{}^\a&= -\frac{1}{120}(D_1^{cdef}
D_1^{c'd'e'f'})_{\m}{}^\alpha G^{(0)}_{cdef}G^{(0)}_{c'd'e'f'}
+\frac{1}{40}\Big(F_1^{ef}+F_2^{ef}+2F_3^{ef} 
\Big)^\a{}_{\m\b}T_{ef}^{(1)}{}^\b~,
\end{alignat}
\bead
E^{(0)}_{~~m}{}^\a&=\Psi_m{}^\a\nn\\
E^{(1)}_{~~m}{}^\a&=\frac{1}{4}(\theta\C^{ef})^\alpha \o_{mef}
-(\theta\cT_m{}^{cdef})^\alpha 
G^{(0)}_{cdef}\nn\\
E^{(2)}_{~~m}{}^\a&=\frac{i}{2}(\Psi_m D_1^{cdef})^\alpha G^{(0)}_{cdef}
+\frac{i}{2}(T_{ef}^{(0)}D_{2m}{}^{ef})^\a\nn\\
E^{(3)}_{~~m}{}^\a&=
\frac{i}{24}(\theta\C^{ab}D_1^{cdef})^\alpha \o_{mab} G^{(0)}_{cdef}
-\frac{i}{6}(\theta\cT_m{}^{cdef}D_1^{c'd'e'f'})^\alpha 
G^{(0)}_{cdef}G^{(0)}_{c'd'e'f'}\nn\\
&+\frac{i}{6}(T_{ef}^{(1)}D_{2m}{}^{ef})^\a
-\frac{1}{6}\Big(F_1^{ef}+F_2^{ef}+2F_3^{ef} 
\Big)^\a{}_{\b\c}\Psi_m{}^\b T_{ef}^{(0)}{}^\c\nn\\ 
E^{(4)}_{~~m}{}^\a&= -\frac{1}{24}(\Psi_m D_1^{cdef}
D_1^{c'd'e'f'})^\alpha G^{(0)}_{cdef}G^{(0)}_{c'd'e'f'}
-\frac{1}{24}(T_{ef}^{(0)}D_{2m}{}^{ef} D_1^{cdgh})^\alpha G^{(0)}_{cdgh}\nn\\
&-\frac{1}{24}\Big(F_1^{ef}+2F_2^{ef}+3F_3^{ef} 
\Big)^\a{}_{\b\c}\Psi_m{}^\b T_{ef}^{(1)}{}^\c
+\frac{i}{12}(T^{(2)}_{ef} D_{2m}{}^{ef})^\a\nn\\
&-\frac{1}{24}
\Big(2F_1^{ef}+F_2^{ef}+3F_3^{ef} 
\Big)^\a{}_{\b\c}
(
\frac{1}{4}\theta\C^{ab}\o_{mab}
-\theta\cT_m{}^{cdgh}
G^{(0)}_{cdgh})^\b 
T_{ef}^{(0)}{}^\c
~.
\end{alignat}

\subsection*{$C$-field expansions}

\bead
C^{(0)}_{\m\n\s}&=0\nn\\
C^{(1)}_{\m\n\s}&=0\nn\\
C^{(2)}_{\m\n\s}&=0\nn\\
C^{(3)}_{\m\n\s}&=\frac{i}{8}(\theta\C_{ab})_{(\m}
(\theta\C^{a})_\n(\theta\C^{b})_{\s)}\nn\\
C^{(4)}_{\m\n\s}&=0
~,
\end{alignat}
\bead
C^{(0)}_{\m\n s}&=0\nn\\
C^{(1)}_{\m\n s}&=0\nn\\
C^{(2)}_{\m\n s}&=\frac{1}{4}(\theta\C_{se})_{(\m}(\theta\C^e)_{\n)}\nn\\
C^{(3)}_{\m\n s}&=\frac{i}{20}(\theta\C^{a})_\m(\theta\C^{b})_{\n}
(\Psi_s\C_{ab}\theta)
+\frac{i}{5}(\theta\C_{ab})_{(\m}(\theta\C^{a})_{\n)}
(\Psi_s\C^{b}\theta)\nn\\
C^{(4)}_{\m\n s}&=\frac{i}{24}
(\theta\C^{a})_\m(\theta\C^{b})_{\n}
\{
-\frac{1}{2}(\th\th)\o_{sab}+\frac{1}{4}(\th\C_{abcd}\th)\o_s{}^{cd}
-(\th\cT_s{}^{cdef}\C_{ab}\th)G^{(0)}_{cdef}
\}\nn\\
&+\frac{i}{12}(\theta\C_{ab})_{(\m}(\theta\C^{a})_{\n)}
\{
\frac{1}{4}(\th\C^{bef}\th)\o_{sef}-(\th\cT_s{}^{cdef}\C^b\th)G^{(0)}_{cdef}
\}\nn\\
&-\frac{i}{72}(D_1^{abcd}\C^e\th)_{(\m|}(\th\C_{se})_{|\n)} G^{(0)}_{abcd}
+\frac{i}{36}(D_1^{abcd}\C_{se}\th)_{(\m}(\th\C^e)_{\n)} G^{(0)}_{abcd}
~,
\end{alignat}

\bead
C^{(0)}_{\s mn}&=0\nn\\
C^{(1)}_{\s mn}&=-\frac{i}{2}(\th\C_{mn})_\s\nn\\
C^{(2)}_{\s mn}&=\frac{2}{3}(\Psi_{[m}\C^e\th)(\th\C_{n]e})_\s
-\frac{1}{3}(\Psi_{[m}\C_{n]e}\th)(\th\C^{e})_\s\nn\\
C^{(3)}_{\s mn}&=\frac{1}{4}
\{
\frac{1}{4}(\th\C^{egh}\th)\o_{[m|gh}-(\th\cT_{[m|}{}^{cdgh}\C^e\th)
G^{(0)}_{cdgh}
\}(\th\C_{|n]e})_\s\nn\\
&-\frac{1}{4}
\{
\frac{1}{4}(\th\C^{gh}{}_{e[m}\th)\o_{n]gh}
-\frac{1}{2}(\th\th)\o_{[mn]e}
-(\th\cT_{[m}{}^{cdgh}\C_{n]e}\th)
G^{(0)}_{cdgh}
\}(\th\C^e)_\s\nn\\
&+\frac{i}{4}(\Psi_m\C^a\th)(\Psi_n\C^b\th)(\th\C_{ab})_\s
-\frac{i}{4}(\Psi_{[m}\C^a\th)(\Psi_{n]}\C_{ab}\th)(\th\C^{b})_\s\nn\\
&+\frac{1}{24}(D_1^{cdgh}\C_{mn}\th)_\s G^{(0)}_{cdgh}\nn\\
C^{(4)}_{\s mn}&=
\frac{i}{15}
\{(\Psi_{[m|} D_1^{cdgh}\C^e\th)G^{(0)}_{cdgh}
+(T^{(0)}_{gh}D_{2[m|}{}^{gh}\C^e\th)
\}(\th\C_{|n]e})_\s\nn\\
&+\frac{i}{5}
\{\frac{1}{4}(\th\C^{agh}\th)\o_{[m|gh}
-(\th\cT_{[m|}{}^{cdgh}\C^a\th)G^{(0)}_{cdgh}
\}(\Psi_{|n]}\C^b\th)(\th\C_{ab})_\s\nn\\
&+\frac{i}{15}(\Psi_{[m}\C^e\th)(D_1^{cdgh}\C_{n]e}\th)_\s G^{(0)}_{cdgh}
-\frac{i}{60}(\Psi_{[m}\C_{n]e}\th)(D_1^{cdgh}\C^{e}\th)_\s G^{(0)}_{cdgh}\nn\\
&-\frac{i}{10}
\{(\Psi_{[m} D_1^{cdgh}\C_{n]e}\th)G^{(0)}_{cdgh}
+(T^{(0)}_{gh}D_{2[m}{}^{gh}\C_{n]e}\th)
\}(\th\C^{e})_\s\nn\\
&+\frac{i}{5}
\{
-\frac{1}{2}(\th\th)\o_{[m|ab}
+\frac{1}{4}(\th\C^{gh}{}_{ab}\th)\o_{[m|gh}
-(\th\cT_{[m|}{}^{cdgh}\C_{ab}\th)G^{(0)}_{cdgh}
\}(\Psi_{|n]}\C^a\th)(\th\C^{b})_\s\nn\\
&-\frac{i}{10}
\{\frac{1}{4}(\th\C^{agh}\th)\o_{[m|gh}
-(\th\cT_{[m|}{}^{cdgh}\C^a\th)G^{(0)}_{cdgh}
\}(\Psi_{|n]}\C_{ab}\th)(\th\C^{b})_\s
~,
\end{alignat}
\bead
4\partial_{[s}C^{(0)}_{mnp]}&=G^{(0)}_{smnp}\nn\\
C^{(1)}_{mnp}&=-3i(\Psi_{[m}\C_{np]}\theta)\nn\\
C^{(2)}_{mnp}
&=-\frac{3i}{2}\{
-\frac{1}{2}(\th\th)\o_{[mnp]}+\frac{1}{4}(\th\C^{gh}{}_{[mn}\th)
\o_{p]gh}-(\th\cT_{[m}{}^{cdgh}\C_{np]}\th)G^{(0)}_{cdgh}
\}\nn\\
&-3(\Psi_{[m}\C^e\theta)(\Psi_n\C_{p]e}\th)\nn\\
C^{(3)}_{mnp}&=\frac{1}{2}(\Psi_{[m}D_1^{cdgh}\C_{np]}\th)G^{(0)}_{cdgh}
+\frac{1}{2}(T^{(0)}_{gh}D_{2[m}{}^{gh}\C_{np]}\th)\nn\\
&+2(\Psi_{[m|}\C^e\th)
\{
\frac{1}{2}(\th\th)\o_{|np]e}-\frac{1}{4}(\th\C^{gh}{}_{e|n}\th)
\o_{p]gh}+(\th\cT_{|n}{}^{cdgh}\C_{p]e}\th)G^{(0)}_{cdgh}
\}\nn\\
&-(\Psi_{[m}\C_{n|}{}^e\th)
\{
\frac{1}{4}(\th\C^{gh}{}_e\th)\o_{|p]gh}-
(\th\cT_{|p]}{}^{cdgh}\C_{e}\th)G^{(0)}_{cdgh}
\}\nn\\
&+i(\Psi_{[m}\C^a\th)(\Psi_{n}\C^b\th)(\Psi_{p]}\C_{ab}\th)\nn\\
C^{(4)}_{mnp}&=
\frac{1}{32}
(\th\C^{gh}D_1^{cdef}\C_{[mn}\th)\o_{p]gh}G^{(0)}_{cdef}
-\frac{1}{8}(\th\cT_{[m}{}^{cdgh}D_1^{c'd'g'h'}\C_{np]}\th)G^{(0)}_{cdgh}
G^{(0)}_{c'd'g'h'}\nn\\
&+\frac{1}{8}(T^{(1)}_{gh}D_{2[m}{}^{gh}\C_{np]}\th)
+\frac{i}{8}(F_1^{gh}+F_2^{gh}+2F_3^{gh})^\a{}_{\b\c}
(\C_{[mn}\th)_\a
\Psi_{p]}{}^\b 
T^{(0)}_{gh}{}^\c \nn\\
&-\frac{3i}{4}
(\Psi_{[m|}\C^e\th)
\{
(\Psi_{|n}D_1^{cdgh}\C_{p]e}\th)G^{(0)}_{cdgh}
+(T^{(0)}_{gh}D_{2|n}{}^{gh}\C_{p]e}\th)
\}\nn\\
&+\frac{3i}{4}
(\Psi_{[m}\C^a\th)(\Psi_{n|}\C^b\th)
\{
-\frac{1}{2}(\th\th)\o_{|p]ab}+
\frac{1}{4}(\th\C^{gh}{}_{ab}\th)\o_{|p]gh}-
(\th\cT_{|p]}{}^{cdgh}\C_{ab}\th)G^{(0)}_{cdgh}
\}\nn\\
&-\frac{i}{4}
(\Psi_{[m}\C_{n|e}\th)
\{
(\Psi_{|p]}D_1^{cdgh}\C^{e}\th)G^{(0)}_{cdgh}
+(T^{(0)}_{gh}D_{2|p]}{}^{gh}\C^{e}\th)
\}\nn\\
&-\frac{3i}{4}
(\Psi_{[m}\C^a\th)(\Psi_{n|}\C_{ab}\th)
\{
\frac{1}{4}(\th\C^{bgh}\th)\o_{|p]gh}-
(\th\cT_{|p]}{}^{cdgh}\C^{b}\th)G^{(0)}_{cdgh}
\}\nn\\
&-\frac{3}{4}
\{
\frac{1}{4}(\th\C^{egh}\th)\o_{[m|gh}-
(\th\cT_{[m|}{}^{cdgh}\C^{e}\th)G^{(0)}_{cdgh}
\}\nn\\
&~~~~~~~~~~~~~~~~~~~~~~\times
\{-\frac{1}{2}(\th\th)\o_{|np]e}+
\frac{1}{4}(\th\C^{g'h'}{}_{e|n}\th)\o_{p]g'h'}-
(\th\cT_{|n}{}^{c'd'g'h'}\C_{p]e}\th)G^{(0)}_{c'd'g'h'}
\}
~.
\end{alignat}


\section{Expansion in the number of fields}

In this section we expand 
our superspace geometry around 
the flat-space solution
\bead
h_m{}^a&:=e_m{}^a-\d_m^a=0\nn\\
\Psi_m{}^\a&=0\nn\\
C^{(0)}_{mnp}&=0 
\label{fields}
\end{alignat}
and we show how to obtain
results exact to 
all orders in the $\theta$-expansion, 
but perturbative 
in the number of fields $h_m{}^a,\Psi_m{}^\a,C^{(0)}_{mnp}$. 
Note that in an expansion around flat space, the connection
$\o_m$ and the covariant field strengths 
$G^{(0)}_{abcd}$, $T^{(0)}_{~~ab}{}^\a$ and $R^{(0)}_{abcd}$ 
start at linear order in the fields (\ref{fields}). More 
precisely, from equations (\ref{2.9}), 
(\ref{four}), (\ref{tors}), (\ref{curv}) we see that
\bead
\o_{nkm}&=\partial_{[k}h_{m]}{}^a\eta_{an}
-\partial_{[m}h_{n]}{}^a\eta_{ak}
-\partial_{[n}h_{k]}{}^a\eta_{am}+\dots\nn\\
G^{(0)}_{abcd}&=4\d_{a}{}^m\d_{b}{}^n\d_{c}{}^p\d_{d}{}^q 
\partial_{[m}C^{(0)}_{npq]}\dots\nn\\
T^{(0)}_{~~ab}{}^\a
&=2\d_{a}{}^m\d_{b}{}^n\partial_{[m}\Psi_{n]}{}^\a+\dots\nn\\
R^{(0)}_{mnpq}&=\d_{a}{}^m\d_{b}{}^n\d_{c}{}^p\d_{d}{}^q
R(\o)_{mnpq}+\dots~,
\end{alignat}
where the ellipses indicate terms quadratic or higher in the fields.

Going back to section 4 we split (\ref{roufa1}) 
as follows
\be
T^{(n)}_{~~ab}{}^\a~ =\frac{i}{n(n-1)} (\cM_{[a|}{}^{ef})^\a{}_\b
 ~\d_{|b]}{}^m\partial_mT^{(n-2)}_{~~~~~ef}{}^\b +Q^{(n)}_{~~ab}{}{}^\a~,
%
\label{kjl}
\end{equation}
where the nonlinear piece $Q$ reads
\bead
Q^{(n)}_{~~ab}{}{}^\a~
&:=\frac{i}{n(n-1)} (\cN_{ab}{}^{c_1\dots c_6})^\a{}_\b(G_{c_1\dots c_4}
T_{c_5c_6}{}^\b )^{(n-2)}\nn\\
&-\frac{i}{n(n-1)} (\cM_{[a|}{}^{ef})^\a{}_\b
 ~h_{|b]}{}^m\partial_mT^{(n-2)}_{~~~~~ef}{}^\b \nn\\
&+\frac{i}{n(n-1)}(\cM_{[a|}{}^{ef})^\a{}_\b
~e_{|b]}{}^m \sum_{r=0}^{n-2}(\O_m^{(r)}T^{(n-r-2)})_{ef}{}^\b
\nn\\
&+\frac{i}{n(n-1)}(\cM_{[a|}{}^{ef})^\a{}_\b
\sum_{r=1}^{n-2}
E^{(r)}_{~~|b]}{}^m(\nabla_m T_{ef}{}^\b)^{(n-r-2)}\nn\\
&+\frac{i}{n(n-1)}(\cM_{[a|}{}^{ef})^\a{}_\b
\sum_{r=0}^{n-2}
E^{(r)}_{~~|b]}{}^\m(\nabla_\m T_{ef}{}^\b)^{(n-r-2)}
~.
\label{coala}
\end{alignat}
We have defined
\be
h_{a}{}^m:=\d_a^n h_{n}{}^b  \d_b^m~,
\label{h1}
\end{equation}
so that
\be
e_{a}{}^m=\d_a^m-h_{a}{}^m+{\cal O}(h^2)~.
\end{equation}
We can rewrite (\ref{kjl}) more compactly using matrix notation and omitting
spinor indices
\be
T^{(n)}_{ab}=\frac{1}{n(n-1)}[\cO]_{ab}{}^{ef}T^{(n-2)}_{ef}+Q^{(n)}_{ab}  ,
\label{cmpct}
\end{equation}
where
\be
[\cO]_{ab}{}^{ef} :=i[\cM_{[a}{}^{ef}] \d_{b]}{}^m\partial_m ~.
\label{yii}
\end{equation}
The point of the split in (\ref{cmpct}) is that once $T^{(n)}$ is evaluated
to any given order in the number of fields, $Q^{(n)}$ can 
be evaluated to the next order. This follows the fact
that $Q^{(n)}$ starts at quadratic order in the number fields, 
as can be seen from definition (\ref{coala}).

Iterating (\ref{cmpct}) we obtain
\bead
T^{(2k)}_{ab}&=\frac{1}{(2k)!}[\cO^{k}]_{ab}{}^{ef}T_{ef}^{(0)}
+\sum_{r=0}^{k-1}
\frac{(2k-2r)!}{(2k)!}[\cO^{r}]_{ab}{}^{ef}Q_{ef}^{(2k-2r)}\nn\\
T^{(2k+1)}_{ab}&=\frac{1}{(2k+1)!}[\cO^{k}]_{ab}{}^{ef}T_{ef}^{(1)}
+\sum_{r=0}^{k-1}
\frac{(2k-2r+1)!}{(2k+1)!}[\cO^{r}]_{ab}{}^{ef}Q_{ef}^{(2k-2r+1)}~,
\label{cmpct1}
\end{alignat}
where $T^{(1)}$ was given explicitly in (\ref{deft1}). 
Using the above results, and the equations of the previous section, 
the vielbein and the connection can be computed iteratively
in the number of fields. The same is true for the inverse
vielbein $E_A{}^M$, which can be seen as follows. Let us assume that
$E_M{}^A$ is known to $k$-th order in the fields and that 
$E_A{}^M$ is known to $(k-1)$-th order. Equation (\ref{4}) implies
\bead
E^{(n)}_{~~b}{}^\mu &=-e_b{}^m E^{(n)}_{~~m}{}^\a\d_\a{}^\mu
-\sum_{r=1}^{n-1}E^{(r)}_{~~b}{}^{m}E^{(n-r)}_{~~~~~m}{}^\a\d_\a{}^\mu\nn\\
&-\sum_{r=0}^{n-1}E^{(r)}_{~~b}{}^{\nu}E^{(n-r)}_{~~~~~\nu}{}^\a\d_\a{}^\mu
+\sum_{r=0}^{n-1}E^{(r)}_{~~b}{}^{M}E^{(n-r)}_{~~~~~M}{}^a\Psi_a{}^\mu~.
\label{invrec1}
\end{alignat}
Clearly, the first term on the right-hand-side is known to $k$-th order. 
The remaining terms on the right-hand-side are also known to $k$-th order, 
because they start at quadratic order in the fields. Hence 
$E^{(n)}_{~~b}{}^\mu$ can be computed to $k$-th order, using 
the formula above. 

Similarly, for $E_{b}{}^m$ we have
\bead
E^{(n)}_{~~b}{}^m &=-e_b{}^p E^{(n)}_{~~p}{}^ae_a{}^m
+\frac{i}{2}E^{(n-1)}_{~~~~~b}{}^{\m}(\C^m\theta)_\mu\nn\\
&-\sum_{r=1}^{n-1}E^{(r)}_{~~b}{}^{p}E^{(n-r)}_{~~~~~p}{}^ae_a{}^m
-\sum_{r=0}^{n-2}E^{(r)}_{~~b}{}^{\mu}E^{(n-r)}_{~~~~~\mu}{}^ae_a{}^m~.
\label{invrec2}
\end{alignat}
Note that 
the first line on the rhs is known to $k$-th order, because 
$E_{p}{}^a$ is so known by assumption and we have shown that 
$E_{b}{}^{\m}$ can be computed to that order. The remaining 
terms are also known to $k$-th order, because 
they start at quadratic order in the fields. One argues 
in the same fashion for the remaining $E_{\beta}{}^{M}$ components 
of the inverse vielbein.

Let us now illustrate the use of the 
formulae above, by computing 
the vielbein exactly in $\theta$ 
and up to quadratic order in the fields.

\subsection{Linear order}

In this case $Q_{ab}=0$ and formula (\ref{cmpct1}) reduces to
\bead
T_{ab}
=[cosh\sqrt{\cO}]_{ab}{}^{ef}T^{(0)}_{ef}
+[\cO^{-1/2}
sinh\sqrt{\cO}]_{ab}{}^{ef}T^{(1)}_{ef}~,
\label{tlin}
\end{alignat}
where the functions of $\cO$ above are formally 
defined by their Taylor expansions around zero.
Of course the series terminate at order $\theta^{32}$.
Note that in the linear approximation $T^{(1)}$ is given by
\be
T^{(1)}_{ef}=
\frac{1}{4}(\theta\C^{cd})^\alpha R^{(0)}_{efcd}
+2(\theta\cT_{[e}{}^{cdgh})^\alpha \d_{f]}{}^m \partial_m G_{cdgh}^{(0)}
~.
\end{equation}
Moreover substituting (\ref{tlin}) 
into (\ref{grtrec}), we arrive at the following 
linear-order expression 
for the four-form
\bead
G_{abcd}
=G^{(0)}_{abcd}+
6i\theta \C_{[ab}\sum_{k=0}\{
\frac{1}{(2k+1)!}\cO^{k}~T^{(0)}
+\frac{1}{(2k+2)!}\cO^{k}~T^{(1)}\}_{cd]}
~.
\label{glin}
\end{alignat}
As a check the reader can verify for herself that
\be
\theta^\mu\partial_\mu G_{abcd}=\theta^\a\nabla_\alpha G_{abcd}
~,
\end{equation}
as follows from (\ref{glin}),(\ref{derivatives}),
(\ref{tlin}), taking into account that 
$\theta^\mu\partial_\mu \cO =2\cO$ and 
$\theta^\mu\partial_\mu T^{(1)}=T^{(1)}$.

Equipped with (\ref{tlin}),(\ref{glin}),  we can now 
systematically derive expressions for all the other 
relevant superspace quantities. 
The strategy is straightforward: truncate 
to linear order the expressions for the superconnection 
and vielbein components given in section 4.2, 
substituting (\ref{tlin}),(\ref{glin}) where necessary. 
For the superconnection components we get
%
%
%
%
%
%
\bead
\O_{\mu ab}
&=\frac{i}{2}(\theta\cR_{ab}{}^{cdef})_\mu G^{(0)}_{cdef}\nn\\
&+\sum_{k=0}\frac{1}{2k+3} \Big(\frac{1}{2k+1}C_{1ab}{}^{ef}
+\frac{1}{2}C_{2ab}{}^{ef}
\Big)_{\mu \a}\{\frac{\cO^{k}}{(2k)!}T^{(0)}\}_{ef}{}^\a\nn\\
&+\sum_{k=0} \frac{1}{2k+4}\Big(\frac{1}{2k+2}C_{1ab}{}^{ef}
+\frac{1}{2}C_{2ab}{}^{ef}
\Big)_{\mu \a}\{\frac{\cO^{k}}{(2k+1)!}T^{(1)}\}_{ef}{}^\a
\end{alignat}
and
\be
\O_{mab}=\o_{mab}+
i\theta \cS_{mab}{}^{ef}\sum_{k=0}\{
\frac{\cO^{k}}{(2k+1)!}T^{(0)}
+\frac{\cO^{k}}{(2k+2)!}T^{(1)}\}_{ef}
~.
\end{equation}
For the vielbein components we get
%
%
%
%
%
%
%
%
\be
E_\m{}^\a=\d_\m^\a+\D E_\m{}^\a~,
\end{equation}
where
\begin{center}
\fbox{\parbox{15.6cm}{
%
\bead
\D E_\m{}^\a&:=\frac{i}{6}(D_1^{abcd})_\m{}^\alpha G^{(0)}_{abcd}\nn\\
&+\sum_{k=0}\frac{1}{2k+4}\Big(\frac{F_1^{ef}}{(2k+3)(2k+1)}
+ \frac{F_2^{ef}}{2(2k+3)}+\frac{F_3^{ef}}{2(2k+1)}\Big)^\a{}_{\mu \b}
\{\frac{\cO^{k}}{(2k)!}T^{(0)}\}_{ef}{}^\b\nn\\
&+\sum_{k=0}\frac{1}{2k+5}\Big(\frac{F_1^{ef}}{(2k+4)(2k+2)} 
+ \frac{F_2^{ef}}{2(2k+4)}+\frac{F_3^{ef}}{2(2k+2)}\Big)^\a{}_{\mu \b}
\{\frac{\cO^{k}}{(2k+1)!}T^{(1)}\}_{ef}{}^\b\nn
\end{alignat}
}}
\end{center}
\be\label{de1}\end{equation}
and
\be
E_m{}^\a=\D E_m{}^\a~,
\end{equation}
\vfill\break
where
\begin{center}
\fbox{\parbox{11.6cm}{
%
\bead
\D E_m{}^\a
&:=\Psi_m{}^\a+\frac{1}{4}(\theta\C^{ab})^\a\o_{mab}
-(\theta\cT_m{}^{abcd})^\alpha G^{(0)}_{abcd}\nn\\
&+i\sum_{k=0}\{\frac{\cO^{k}}{(2k+2)!}
T^{(0)}
+\frac{\cO^{k}}{(2k+3)!}T^{(1)}
\}_{ef}{}^\b(D_{2m}{}^{ef})_\b{}^\alpha ~.\nn
\end{alignat}
}}
\end{center}
\be\label{de2}\end{equation}
Finally, equations (\ref{rec1}),(\ref{rec2}) imply
\be
E_\m{}^a=-\frac{i}{2}(\C^a\theta)_\m+ \D E_\m{}^a~,
\end{equation}
where
\begin{center}
\fbox{\parbox{15.6cm}{
%
\bead
\D E_\m&{}^a:=\frac{1}{24}
(D_1^{bcde}\C^a\theta)_\m G^{(0)}_{bcde}\nn\\
-&\sum_{k=0}\frac{i(\C^a\theta)_\a}
{(2k+5)(2k+4)}\Big(\frac{F_1^{ef}}{(2k+3)(2k+1)}
+ \frac{F_2^{ef}}{2(2k+3)}+\frac{F_3^{ef}}{2(2k+1)}\Big)^\a{}_{\mu \b}
\{\frac{\cO^{k}}{(2k)!}T^{(0)}\}_{ef}{}^\b\nn\\
-&\sum_{k=0}\frac{i(\C^a\theta)_\a}
{(2k+6)(2k+5)}\Big(\frac{F_1^{ef}}{(2k+4)(2k+2)} 
+ \frac{F_2^{ef}}{2(2k+4)}+\frac{F_3^{ef}}{2(2k+2)}\Big)^\a{}_{\mu \b}
\{\frac{\cO^{k}}{(2k+1)!}T^{(1)}\}_{ef}{}^\b\nn
\end{alignat}
}}
\end{center}
\be\label{de3}\end{equation}
and
\be
E_m{}^a=\d_m{}^a+\D E_m{}^a~,
\end{equation}
where
\begin{center}
\fbox{\parbox{11.6cm}{
%
\bead
\D E_m{}^a
&:=h_m{}^a-i(\Psi_m \C^a\theta)
-\frac{i}{8}(\theta\C^{aef}\theta)~\o_{mef}
+\frac{i}{2}(\theta\cT_m{}^{bcde}\C^a\theta)~G^{(0)}_{bcde}\nn\\
&+\sum_{k=0}\{\frac{\cO^{k}}{(2k+3)!}
T^{(0)}
+\frac{\cO^{k}}{(2k+4)!}T^{(1)}
\}_{ef}{}^\b(D_{2m}{}^{ef}\C^a\theta)_\b ~.\nn
\end{alignat}
}}
\end{center}
\be\label{de4}\end{equation}
We will also need the inverse vielbein components $E_b{}^M$ to 
linear order. From equations (\ref{invrec1}), (\ref{invrec2}) we obtain
\bead
E_b{}^\m
&=-\Psi_b{}^\m-\frac{1}{4}(\theta\C^{ef})^\m\o_{bef}
+(\theta\cT_b{}^{cdef})^\m G^{(0)}_{cdef}\nn\\
&-i\sum_{k=0}\{\frac{\cO^{k}}{(2k+2)!}
T^{(0)}
+\frac{\cO^{k}}{(2k+3)!}T^{(1)}
\}_{ef}{}^\b(D_{2b}{}^{ef})_\b{}^\m 
\end{alignat}
and
\bead
E_b{}^m
&=e_b{}^m+\frac{i}{2}(\Psi_b\C^m\theta)\nn\\
&+\sum_{k=0}\{ \frac{2k+1}{2}\frac{\cO^{k}}{(2k+3)!}
T^{(0)}
+\frac{2k+2}{2}\frac{\cO^{k}}{(2k+4)!}T^{(1)}
\}_{ef}{}^\b(D_{2b}{}^{ef}\C^m\theta)_\b ~.
\end{alignat}
It is straightforward to check that to linear order the 
expressions above satisfy $E_b{}^M E_{M}{}^A=\d_{b}{}^A$, 
as they should.

For the $C$-field we get

\be
C_{\m\n\s}=\frac{i}{8}(\C^a\th)_{(\m}(\C^b\th)_\n(\C_{ab}\th)_{\s)}
+\D C_{\m\n\s}~,
\end{equation}
where
\be
\D C_{\m\n\s}:=\sum_{n=0}\{
\frac{3i}{4(n+6)}(\C^a\th)_{(\m}(\C^b\th)_\n
\D E^{(n)}_{~~\s)}{}^\a(\C_{ab}\th)_{\a}
-\frac{3}{n+5}\D E^{(n)}_{~~(\m}{}^a(\C^b\th)_\n(\C_{ab}\th)_{\s)}
\}~.
\label{dc1}
\end{equation}
\be
C_{s\m\n}=\frac{1}{4}(\C^a\th)_{(\m}(\C_{ab}\th)_{\n)}\d_s^b
+\D C_{s\m\n}~,
\end{equation}
where
\bead
\D C_{s\m\n}:=\sum_{n=0}\{&
\frac{i}{4(n+5)}(\C^a\th)_{\m}(\C^b\th)_\n
\D E^{(n)}_{~~s}{}^\a(\C_{ab}\th)_{\a}
-\frac{1}{n+4}\D E^{(n)}_{~~s}{}^a(\C^b\th)_{(\m}(\C_{ab}\th)_{\n)}\nn\\
&+\frac{2i}{n+3}\D E^{(n)}_{~~(\m}{}^a(\C_{as}\th)_{\n)}
+\frac{1}{n+4}\D E^{(n)}_{~~(\m}{}^\a(\C_{sa}\th)_{\a}(\C^a\th)_{\n)}
\}~.
\label{dc2}
\end{alignat}
\be
C_{mn\s}=-\frac{i}{2}(\C_{ab}\th)_{\s}\d_m^a\d_n^b
+\D C_{mn\s}~,
\end{equation}
where
\bead
\D C_{mn\s}:=
\sum_{n=0}\{&
\frac{2i}{n+2}\D E^{(n)}_{~~[m}{}^a(\C_{n]a}\th)_{\s}
-\frac{i}{n+2}\D E^{(n)}_{~~\s}{}^\a(\C_{mn}\th)_{\a}\nn\\
&-\frac{1}{n+3}\D E^{(n)}_{~~[m}{}^\a(\C_{n]a}\th)_{\a}(\C^a\th)_{\s}
\}~.
\label{dc3}
\end{alignat}
\be
C_{mnp}=\D C_{mnp}~,
\end{equation}
where
\bead
\D C_{mnp}:=C^{(0)}_{mnp}-\sum_{n=0}
\frac{3i}{n+1}\D E^{(n)}_{~~[m}{}^\a(\C_{np]}\th)_{\a}
~.
\label{dc4}
\end{alignat}

\subsection{Quadratic order and beyond.}

In order to compute $T_{ab}{}^\a$ to quadratic order, we 
have to determine $Q_{ab}{}^\a$. Ignoring terms 
that start at cubic order, from equation (\ref{coala}) we have
\bead
Q^{(n)}_{~~ab}{}{}^\a~
&=\frac{i}{n(n-1)} (\cN_{ab}{}^{c_1\dots c_6})^\a{}_\b(G_{c_1\dots c_4}
T_{c_5c_6}{}^\b )^{(n-2)}\nn\\
&-\frac{i}{n(n-1)} (\cM_{[a|}{}^{ef})^\a{}_\b
 ~h_{|b]}{}^m\partial_mT^{(n-2)}_{~~~~~ef}{}^\b \nn\\
&+\frac{i}{n(n-1)}(\cM_{[a|}{}^{ef})^\a{}_\b
~\d_{|b]}{}^m \sum_{r=0}^{n-2}(\O_m^{(r)}T^{(n-r-2)})_{ef}{}^\b
\nn\\
&+\frac{i}{n(n-1)}(\cM_{[a|}{}^{ef})^\a{}_\b
\sum_{r=1}^{n-2}
E^{(r)}_{~~|b]}{}^m \partial_m T^{(n-r-2)}_{~~~~~~~~ef}{}^\b\nn\\
&+\frac{i}{n(n-1)}(\cM_{[a|}{}^{ef})^\a{}_\b
\sum_{r=0}^{n-2}
E^{(r)}_{~~|b]}{}^\m\partial_\m T^{(n-r-1)}_{~~~~~~~~ef}{}^\b
~.
\label{above}
\end{alignat}
Therefore $Q_{ab}{}^\a$ can be readily 
obtained by plugging into (\ref{above})
the {\it linear} 
expressions for $G$, $T$, $\O_m$, $E_b{}^M$ obtained in 
section 6.1. Then $T$ is computed 
to quadratic order using equation (\ref{cmpct1}) 
and $G$ is computed 
to quadratic order through the recursion 
relation (\ref{grtrec}). 
The connection and the vielbein are obtained 
to quadratic order by using formulae 
(\ref{rec1}),(\ref{rec2}),(\ref{rec3}),(\ref{omrec}),(\ref{rec42}),(\ref{rec52}). 
Clearly, this procedure can be iterated to any desired order 
in the number of fields.


\section{The supermembrane}

The eleven-dimensional 
supermembrane can be described by a superembedding 
\be
Z:\S^{(3|0)}\rightarrow M^{(11|32)}~,
\end{equation}
where
\be
Z^{\unM}:=(X^{\unm},\th^{\unmu})
\end{equation}
are the membrane supercoordinates. 
For the purposes of this section we pass to superembedding notation 
whereby all target-space (i.e. eleven-dimensional) indices are 
underlined. The world-volume theory of the membrane 
is given by \cite{bsta, bstb}
\be
S=\int_\S{d\s^3}\{ \sqrt{-g}+f^*C\}~,
\label{smodel}
\end{equation}
where
\be
f^*C:=
\frac{1}{6}\varepsilon^{mnp}\partial_mZ^{\unP}
\partial_nZ^{\unN}\partial_pZ^{\unM} C_{\unM\unN\unP} 
\end{equation}
is the pull back of the $C$-field onto the membrane world-volume 
and $g$ is the determinant of the Green-Schwarz metric
\be
g_{mn}:=\big(\partial_mZ^{\unM}E_{\unM}{}^{\una}\big)
\big(\partial_nZ^{\unN}E_{\unN}{}^{\unb}\big)\eta_{\una\unb}~.
\label{gs}
\end{equation}

\subsection{Linear background}

Using the 
results of section 6, 
we are now ready compute the linear coupling of the supermembrane 
to an on-shell eleven-dimensional background. 
For the metric (\ref{gs}) we find, to linear order,
\be
g_{mn}=G_{mn}+\D g_{mn},
\end{equation}
where
\be
G_{mn}:=\Pi_m{}^{\una} \Pi_n{}^{\unb}\eta_{\una\unb}~,
\end{equation}
\be
\D g_{mn}:=
2\Pi_{(m}{}^{\una} \partial_{n)}Z^{\unN}\D E_{\unN}{}^{\unb}~\eta_{\una\unb}
\end{equation}
and
\be
\Pi_m{}^{\una}:=\partial_m X^{\una}-\frac{i}{2}(\partial_m\th\C^{\una}\th)~;
~~~~~~~X^{\una}:=X^{\unm}\d_{\unm}^{\una}~.
\end{equation}
Note that $G_{mn}$ is the Green-Schwarz metric for the case
of flat target space. The linear-order components 
($\D E_{\unM}{}^{\una}$) of the eleven-dimensional vielbein 
were given explicitly in section 6.1, 
equations (\ref{de1}),(\ref{de2}),(\ref{de3}),(\ref{de4}). 
The determinant is given in the linear approximation by
\be
\sqrt{-g}=\sqrt{-G}~(1+\D g_{mn}G^{mn})~,
\end{equation}
where $G^{mn}$ is the inverse of $G_{mn}$.
Moreover, for the Wess-Zumino term in (\ref{smodel}) we have
\bead
f^*C&=f^*\D C+
\varepsilon^{mnp}\{\frac{i}{4}\partial_mX^{\una}\partial_nX^{\unb}
(\partial_p\th\C_{\una\unb}\th)\nn\\
&+\frac{1}{8}\partial_mX^{\una}(\partial_n\th\C^{\unb}\th)
(\partial_p\th\C_{\una\unb}\th)
-\frac{i}{48}(\partial_m\th\C^{\una}\th)
(\partial_n\th\C^{\unb}\th)(\partial_p\th\C_{\una\unb}\th)  \}~,
\end{alignat}
where the linear-order components ($\D C_{\unM\unN\unP}$) of the 
three-form potential were given explicitly in section 6.1, 
equations (\ref{dc1}),(\ref{dc2}),(\ref{dc3}),(\ref{dc4}). 

To summarize, to linear order the coupling of the supermembrane 
to the eleven-dimensional background is given by
\be
S=S_{flat}+\int_\S{d\s^3}\{ \sqrt{-G}G^{mn}\D g_{mn}+f^*\D C\}~,
\label{lm2}
\end{equation}
where
\bead
S_{flat}:=\int_\S{d\s^3}\{ \sqrt{-G}
+\varepsilon^{mnp}\Big[\frac{i}{4}\partial_mX^{\una}\partial_nX^{\unb}
(\partial_p\th\C_{\una\unb}\th)
&+\frac{1}{8}\partial_mX^{\una}(\partial_n\th\C^{\unb}\th)
(\partial_p\th\C_{\una\unb}\th)\nn\\
&-\frac{i}{48}(\partial_m\th\C^{\una}\th)
(\partial_n\th\C^{\unb}\th)(\partial_p\th\C_{\una\unb}\th)  \big]\}~
\label{flat}
\end{alignat}
is the action of a supermembrane in flat eleven-dimensional target space
\footnote{In order to translate the action (\ref{flat}) to the 
one given in equation (2.1) of \cite{whn}, see footnote 4.}.


\section{Outlook}

Our result in section 7.1 can be used to read off the covariant 
membrane vertex operator, to linear order in the background fields. 
The covariant superstring and superparticle vertex operators 
in the linearized approximation, can also be obtained by 
reduction. One could also use the results of section 6.2 to 
go beyond the linear approximation, to any desired order in the 
number of background fields. This may be a promising approach, 
alternative to \cite{ba}, to 
formulating covariant scattering amplitudes for the superstring 
or the superparticle.

By gauge-fixing to the light-cone gauge, 
one could of course compare with the 
existing results in the literature. The linearized 
light-cone membrane vertex operator of \cite{dnp} terminates 
at order $\th^5$. It would be interesting to know if 
similar simplifications occur even when one introduces 
terms nonlinear in the background fields. In this case, 
one may hope that 
the full $\th$-expansion of the light-cone vertex operator, 
in a general supergravity background, may be computable.

As already mentioned, 
one could regularize (\ref{lm2}) 
in order to make contact with the matrix model and compare with 
the partial results obtained in \cite{tra}. 
Moreover, our methods make it 
readily possible to go beyond the linear order in the 
background fields. Subtleties arise already at the quadratic order 
\cite{dos, do} and it would be of interest to have 
an explicit matrix-model candidate in this case.

Another possible application 
concerns membrane (and also five-brane) 
instanton computations. 
These have so far been 
restricted to the case of membranes wrapping 
rigid, isolated supersymmetric cycles. In order to study the contribution 
of instantons with more zero modes, it is necessary to have 
knowledge of the higher terms in the $\th$-expansion of the 
supermembrane action. The explicit results of section 5 can be used 
for this purpose. 

Finally let us note that the methods of this paper can be readily 
carried over to other superspaces in lower dimensions, 
as well as to the case of deformed eleven-dimensional supergravity, 
where the latter is obtained by modifying the 
torsion constraints as in \cite{sugraa, sugrab}. 
This modification will result in deforming the 
action of the supercovariant spinor derivative --and, therefore, 
the recursion 
relations (\ref{grtrec})-- by the inclusion of higher-order curvature 
terms. In its turn this 
will induce higher-order curvature corrections to the 
remaining recursion relations presented in section 4. In 
\cite{sugraa, sugrab} the deformations to the 
torsion constraints were parametrized in terms of 
certain superfields which were treated as `black boxes'. In order 
to obtain explicit expressions however, these
superfields would ultimately have to be 
expressed in terms of the fields in the massless multiplet. 
Unfortunately at present this remains a very difficult problem,  
although a systematic way to arrive at these explicit corrections 
has been proposed in \cite{ht}.


\section*{Acknowledgments}

We would like to thank A.~Brandhuber, 
G.~Ferretti, P.~Howe, B.~Nilsson, S.~Ramgoolam 
and P.~Vanhove for 
useful discussions. This work is supported by 
EU contract HPRN-CT-2000-00122.

\vfill\break


\section*{Appendix A: Definitions}

For the convenience of the reader in this section we include an 
alphabetical index of various definitions used in this paper.

$(C_{1ab}{}^{ef})_{\a\b }$, 
$(C_{2 ab}{}^{ef})_{\a\b }$: eqn (\ref{cdefs})

$(D_1^{cdef})_\b{}^\a$: eqn (\ref{dfdefs})

$(D_{2a}{}^{bc})_\b{}^\a$: eqn (\ref{ddef})

$\cD_m$: eqn (\ref{conb})

$\D C_{MNP}$: eqns (\ref{dc1}),(\ref{dc2}),(\ref{dc3}),(\ref{dc4})

$\D E_M{}^A$: eqns (\ref{de1}),(\ref{de2}),(\ref{de3}),(\ref{de4})

$e_m{}^a$: eqn (\ref{koko})

$(F_1^{ef})^\a{}_{\b\c}$, 
$(F_2^{ef})^\a{}_{\b\c}$, 
$(F_3^{ef})^\a{}_{\b\c}$: eqn (\ref{dfdefs})

$\C_m$: eqn (\ref{yuu})

$h_m{}^a$: eqn (\ref{fields})

$h_a{}^m$: eqn (\ref{h1})

$(\cM_{a}{}^{ef})^\a{}_\b$:  eqn (\ref{roufa2})

$(\cN_{ab}{}^{c_1\dots c_6})^\a{}_\b$: eqn (\ref{roufa2})

$[\cO]_{ab}{}^{ef}$: eqn (\ref{yii})

$\o_{mA}{}^{B}$: eqn (\ref{koko})

$\Psi_m{}^\a$: eqn (\ref{koko})

$\Psi_a{}^\m$: eqn (\ref{3})

$Q^{(n)}_{~~ab}{}{}^\a$: eqn (\ref{coala})

$\cR_{ab}{}^{cdef}$: eqn (\ref{ugm})

$S^{(n)}_{\{ A\}}$: eqn (\ref{gotyab})

$\cS_{bcd}{}^{ef}$: eqn (\ref{ugm})

$\cT_a{}^{bcde}$: eqn (\ref{ugm})

$\th^\a$: eqn (\ref{kalos})


\section*{Appendix B: Gamma matrices}

In this section we explain our spinor notation 
and conventions for gamma matrices.

Spinor indices are denoted by 
lower-case Greek letters and spinors always carry 
the spinor index ``upstairs''. 
In eleven dimensions there is an antisymmetric 
charge-conjugation matrix, 
\be
C_{\a\b}=-C_{\b\a}~, 
\end{equation}
which can be used to raise/lower indices on gamma matrices:
\be
(\C_a)_{\a\b}:=C_{\a\c}(\C_a)^\c{}_\b; ~~~~~~~
(\C_a)^{\a\b}:=C^{\a\c}(\C_a)_\c{}{}^\b~,
\end{equation}
where $C^{\a\b}$ is the inverse of $C_{\a\b}$.

We denote by $\C^{a_1\dots a_n}$ the antisymmetrized product 
of $n$ gamma matrices
\be
\C^{a_1\dots a_n}:=\C^{[a_1}\dots \C^{a_n]}~.
\end{equation}
For $n=1,2,5$ the antisymmetrized products are symmetric 
in the spinor indices, whereas for $n=0,3,4$ they are antisymmetric. 
In particular we have
\be
\C^a_{\a\b}=\C^a_{\b\a}~.
\end{equation}
In our conventions the Majorana condition reads
\be
\bar{\th}=\th^{Tr}C~.
\end{equation}
We can use the above to drop the bar on the 
fermionic coordinate $\th$. For example,
\be
(\bar{\th}\C_a\l):=(\bar{\th}\C_a)_\a\l^\a=(\th^{Tr}C)_\b(\C_a)^\b{}_\a\l^\a
=\th^\c C_{\c\b}(\C_a)^\b{}_\a\l^\a=
\th^\c (\C_a)_{\c\a}\l^\alpha =(\th\C_a\l)~.
\end{equation}
%


\section*{Appendix C: Pure geometry}

The expansions of section 5 simplify considerably in the case of 
a purely geometric background. By that we mean a bosonic background in which 
the four-form field strength vanishes
\footnote{There is a 
potential global 
obstruction to the vanishing of the four-form 
\cite{w}. We are ignoring such subtleties here.}. 
Of course in this case the on-shell equations of motion 
impose the Ricci-flatness condition.  
Note that $C^{(0)}_{mnp}$ is not necessarily zero. 
This situation appears in many physically interesting 
settings, see for example \cite{hm}. 

In the case of a purely geometric background we have
\bead
T^{(1)}_{~~ab}{}^\a&=\frac{1}{4}(\theta\C^{cd})^\alpha R^{(0)}_{abcd}
\end{alignat}
and
\be
\Psi_m{}^\a=G^{(0)}_{abcd}=T^{(0)}_{~~ab}{}^\a=T^{(2)}_{~~ab}{}^\a=0~.
\end{equation}
Taking the above into account we find
\bead
E^{(0)}_{~~\m}{}^a&=0\nn\\
E^{(1)}_{~~\m}{}^a&=-\frac{i}{2}(\C^a\theta)_\m\nn\\
E^{(2)}_{~~\m}{}^a&=0\nn\\
E^{(3)}_{~~\m}{}^a&=0\nn\\
E^{(4)}_{~~\m}{}^a&=0\nn\\
E^{(5)}_{~~\m}{}^a&=\frac{i}{240}\Big(F_1^{ef}+F_2^{ef}+2F_3^{ef} 
\Big)^\a{}_{\m\b}(\C^a\theta)_\alpha T_{ef}^{(1)}{}^\b~,
\end{alignat}
\bead
E^{(0)}_{~~m}{}^a&=e_m{}^a\nn\\
E^{(1)}_{~~m}{}^a&=0\nn\\
E^{(2)}_{~~m}{}^a&=-\frac{i}{8}(\theta\C^{aef}\theta)\o_{mef}\nn\\
E^{(3)}_{~~m}{}^a&=0\nn\\
E^{(4)}_{~~m}{}^a&=
\frac{1}{24}(T_{ef}^{(1)}D_{2m}{}^{ef}\C^a\theta)\nn\\
E^{(5)}_{~~m}{}^a&=0
\end{alignat}
and
\bead
E^{(0)}_{~~\m}{}^\a&=\d_\m{}^\a\nn\\
E^{(1)}_{~~\m}{}^\a&=0\nn\\
E^{(2)}_{~~\m}{}^\a&=0\nn\\
E^{(3)}_{~~\m}{}^\a&=0\nn\\
E^{(4)}_{~~\m}{}^\a&= \frac{1}{40}\Big(F_1^{ef}+F_2^{ef}+2F_3^{ef} 
\Big)^\a{}_{\m\b}T_{ef}^{(1)}{}^\b~,
\end{alignat}
\bead
E^{(0)}_{~~m}{}^\a&=0\nn\\
E^{(1)}_{~~m}{}^\a&=\frac{1}{4}(\theta\C^{ef})^\alpha \o_{mef}
\nn\\
E^{(2)}_{~~m}{}^\a&=0\nn\\
E^{(3)}_{~~m}{}^\a&=
\frac{i}{6}(T_{ef}^{(1)}D_{2m}{}^{ef})^\a\nn\\ 
E^{(4)}_{~~m}{}^\a&=0
~.
\end{alignat}

\vfill\break

Expansions for the $C$ field:

\bead
C^{(0)}_{\m\n\s}&=0\nn\\
C^{(1)}_{\m\n\s}&=0\nn\\
C^{(2)}_{\m\n\s}&=0\nn\\
C^{(3)}_{\m\n\s}&=\frac{i}{8}(\theta\C_{ab})_{(\m}
(\theta\C^{a})_\n(\theta\C^{b})_{\s)}\nn\\
C^{(4)}_{\m\n\s}&=0
~,
\end{alignat}
\bead
C^{(0)}_{\m\n s}&=0\nn\\
C^{(1)}_{\m\n s}&=0\nn\\
C^{(2)}_{\m\n s}&=\frac{1}{4}(\theta\C_{se})_{(\m}(\theta\C^e)_{\n)}\nn\\
C^{(3)}_{\m\n s}&=0\nn\\
C^{(4)}_{\m\n s}&=-\frac{i}{48}
(\theta\C^{a})_\m(\theta\C^{b})_{\n}
\{
(\th\th)\o_{sab}-\frac{1}{2}(\th\C_{abcd}\th)\o_s{}^{cd}
\}\nn\\
&+\frac{i}{48}(\theta\C_{ab})_{(\m}(\theta\C^{a})_{\n)}
(\th\C^{bef}\th)\o_{sef}~,
\end{alignat}

\bead
C^{(0)}_{\s mn}&=0\nn\\
C^{(1)}_{\s mn}&=-\frac{i}{2}(\th\C_{mn})_\s\nn\\
C^{(2)}_{\s mn}&=0\nn\\
C^{(3)}_{\s mn}&=\frac{1}{16}\o_{[m|gh}
(\th\C^{ghe}\th)(\th\C_{|n]e})_\s\nn\\
&+\frac{1}{8}
\{(\th\th)\o_{[mn]e}-
\frac{1}{2}(\th\C^{gh}{}_{e[m}\th)\o_{n]gh}
\}(\th\C^e)_\s\nn\\
C^{(4)}_{\s mn}&=0
~,
\end{alignat}
\bead
\partial_{[s}C^{(0)}_{mnp]}&=0\nn\\
C^{(1)}_{mnp}&=0\nn\\
C^{(2)}_{mnp}
&=\frac{3i}{4}\{
(\th\th)\o_{[mnp]}-\frac{1}{2}(\th\C^{gh}{}_{[mn}\th)
\o_{p]gh}\}\nn\\
C^{(3)}_{mnp}&=0\nn\\
C^{(4)}_{mnp}&=
+\frac{1}{8}(T^{(1)}_{gh}D_{2[m}{}^{gh}\C_{np]}\th)
\nn\\
&+\frac{3}{32}
(\th\C^{ghe}\th)\o_{[m|gh}
\{(\th\th)\o_{|np]e}-
\frac{1}{2}(\th\C^{g'h'}{}_{e|n}\th)\o_{p]g'h'}
\}
~.
\end{alignat}
\vfill\break


\end{document}